\documentclass{article}
\usepackage{cite}
\usepackage{amsfonts}
\usepackage{amssymb}
\usepackage{amsmath}
\usepackage{graphicx}
\usepackage{slashed}
\usepackage{setspace}

\newcommand{\reef}[1]{(\ref{#1})}

\newcommand{\dash}{\text{-}}
\newcommand{\pdan}{{m}}
\newcommand{\HEW}{{A}}

\newcommand{\ca}{{\cal A}}

\newcommand{\cn}{{\cal N}}
\newcommand{\cm}{{\cal M}}

\newcommand{\cs}{{\cal S}}

\newcommand{\be}{\begin{equation}}
\newcommand{\ee}{\end{equation}}

\def\be{\begin{equation}}
\def\ee{\end{equation}}
\def\bea{\begin{eqnarray}}
\def\eea{\end{eqnarray}}
\def\ba{\begin{array}}
\def\ea{\end{array}}
\def\bd{\begin{displaymath}}
\def\ed{\end{displaymath}}

\def\ie{{\it i.e.~}}

\def\a{\alpha}

\def\d{\delta}
\def\e{\epsilon}           

\def\h{\eta}

\def\l{\lambda}

\def\s{\sigma}                                   

\def\pa{\partial}                              
\def\>{\rangle} 
\def\<{\langle} 
\def\Dsl{D \hskip-.6em \raise1pt\hbox{$ / $ } }
\def\to{\rightarrow}

\def\pa{\partial}

\def\lab{\label}

\newcommand{\eps}{\epsilon}
\newcommand{\lra}{\leftrightarrow}

\def\tQ{\tilde{Q}}

\begin{document}

\setstretch{1.05}

\begin{titlepage}

\begin{flushright}
MIT-CTP-4086 \\
PUPT-2323 \\
MCTP-09-53 \\
\end{flushright}
\vspace{1cm}

\begin{center}
{\Large\bf Solution to the Ward Identities for Superamplitudes} \\[2.5mm]
\vspace{1cm}
{\bf Henriette Elvang${}^{a}$,
Daniel Z.~Freedman${}^{b,c}$, Michael Kiermaier$^{d}$} \\
\vspace{0.7cm}
{{${}^{a}${\it Michigan Center for Theoretical Physics}\\
{\it Randall Laboratory of Physics}\\
{\it University of Michigan}\\
{\it 450 Church St, Ann Arbor, MI 48109, USA}}\\[5mm]
{${}^{b}${\it Center for Theoretical Physics}}\\
{${}^{c}${\it Department of Mathematics}}\\
         {\it Massachusetts Institute of Technology}\\
         {\it 77 Massachusetts Avenue}\\
         {\it Cambridge, MA 02139, USA}}\\[5mm]
{${}^{d}${\it Joseph Henry Laboratories}\\
{\it Princeton University}\\
{\it Princeton, NJ 08544, USA}}\\[5mm]
{\small \tt  elvang@umich.edu,
 dzf@math.mit.edu, mkiermai@princeton.edu}
\end{center}
\vskip .3truecm

\begin{abstract}
\noindent

Supersymmetry and R-symmetry Ward identities relate on-shell amplitudes  in a supersymmetric field theory. We solve these  Ward identities for N$^K$MHV amplitudes of the maximally supersymmetric $\cn=4$ and $\cn=8$ theories. The resulting  superamplitude is written in a new, manifestly supersymmetric and R-invariant form: it is expressed  as a sum of very simple SUSY and  $SU(\cn)_ R$-invariant Grassmann polynomials, each multiplied by a ``basis amplitude". For N$^K$MHV $n$-point superamplitudes  the number of basis amplitudes  is equal to the dimension of the irreducible representation of $SU(n-4)$ corresponding to the rectangular Young diagram with $\cn$ columns and $K$ rows. The linearly independent amplitudes in this algebraic basis may still be functionally related by permutation of momenta. We show how cyclic and reflection symmetries can be used to obtain a smaller functional basis of color-ordered single-trace amplitudes in $\cn=4$ gauge theory. We also analyze the more significant reduction that occurs in $\cn=8$ supergravity because gravity amplitudes are not ordered. All results are valid at both tree and loop level.

\end{abstract}

\end{titlepage}

\setstretch{0.3}
\setcounter{tocdepth}{2}
\tableofcontents
\setstretch{1.05}
\newpage

\setcounter{equation}{0}
\section{Introduction and Summary}

Supersymmetry and R-symmetry Ward identities impose linear relations among the on-shell amplitudes
 of theories with supersymmetry \cite{gris} (see also \cite{gpvanN, PTsusy}).
In $\cn=4$ super Yang-Mills theory (SYM) and in $\cn=8$ supergravity, superamplitudes  $\ca_n$ encode all individual $n$-point amplitudes at each N$^K$MHV level.  The Ward identities can be elegantly and compactly imposed as constraints on the superamplitudes.
The purpose of this paper is to solve these constraints and derive representations in which superamplitudes are expressed as sums of simple
manifestly
SUSY- and
R-invariant Grassmann polynomials, each multiplied by an ordinary ``basis amplitude". We focus on the Ward identities of Poincar\'e SUSY and $SU(\cn)_R$ symmetry,\footnote{Rather than those
of conventional or dual conformal symmetries \cite{Drummond:2006rz,Alday:2007hr,Alday:2007he,Drummond:2008vq,Brandhuber:2008pf,DH,Drummond:2008bq,Drummond:2009fd,Brandhuber:2009kh,Brandhuber:2009xz,Elvang:2009ya,Bargheer:2009qu,Korchemsky:2009hm}.
}
 and our results therefore apply both to tree and loop amplitudes, to both planar and non-planar contributions, and
to both $\cn=4$ SYM theory and $\cn=8$ supergravity.

In the MHV sectors of $\cn=4$ SYM theory and $\cn=8$ supergravity, the SUSY Ward identities imply that any MHV amplitude is proportional to the pure gluon (graviton) amplitude $A_n(++ \dots + -- )$. Beyond the MHV level, however, the SUSY Ward identities give an intricate system of coupled linear equations relating the amplitudes in each N$^K$MHV sector.  The rank of this system tells us how many $n$-point
``basis amplitudes"
 one needs to specify in order to know all $n$-point N$^K$MHV amplitudes.

As an example, consider the NMHV sector of 6-point amplitudes in $\cn=8$ supergravity. There are 151 ways to select 6 external states from
the five different types of particles (gravitons $h$, gravitinos $\psi$, graviphotons $v$, graviphotinos $\chi$ and scalars $\phi$).
Each selection  gives rise to several different amplitudes  due to the multiplicities  $(1,8,28,56,70)$
 of particle states. $SU(8)_R$ symmetry relates
many of them, leaving 1732 independent amplitudes.
However,  our study of the SUSY Ward identities shows that  they are all
 determined by  just 5 functions, namely the basis amplitudes
\bea
 \nonumber
   &&M_{6}(  -+ ++--)\, , ~~~~~~~~
   M_{6}( \psi^-  \psi^+ ++--)\, ,~~~~~~~~
    M_{6}( v^-  v^+ ++--)\, ,~~~~~\\[1mm]
  && M_{6}( \chi^-  \chi^+ ++--)\, ,~~~~~
   M_{6}( \phi^{1234}  \phi^{5678} ++--)\, .
\lab{NMHVba}
\eea
At any loop order, these are the five amplitudes one needs to calculate in order to determine all other NMHV 6-point amplitudes.
This is relevant, for example, in explicit tests of finiteness of $\cn=8$ supergravity for 6-point amplitudes.

Superamplitudes \cite{nair,witten, khoze,BEF} are Grassmann polynomials of degree $\cn(K+2)$  at N$^K$MHV level.
Their coefficients are the actual scattering amplitudes. At tree level, explicit expressions for N$^K$MHV superamplitudes are available for $\cn=4$ SYM theory~[\citen{nair,witten,khoze,BEF},\citen{Brandhuber:2008pf,ArkaniHamed:2008gz,EFK1,DH,EFK2,KN}], and at MHV and NMHV level for $\cn=8$ supergravity~[\citen{witten,BEF,ArkaniHamed:2008gz,Drummond:2009ge}]. Beyond tree level, however, only partial results are available in both theories,\footnote{
The ``link-representation'' for $\cn=4$ SYM
amplitudes~\cite{ArkaniHamed:2009dn} is conjectured to contain
leading-singularity information about amplitudes at any loop order. See also \cite{linkrefs}.}
  and it is therefore worthwhile to systematically explore the general structure of superamplitudes.
Supersymmetry requires that superamplitudes are annihilated by the supercharges, $Q \ca_n = \tQ \ca_n = 0$. The resulting non-trivial constraints on the individual component amplitudes are precisely the SUSY Ward identities. $SU(\cn)_R$-symmetry also plays a central role. The R-symmetry Ward identities, $\delta_R \ca_n = 0$,
 impose further relations among amplitudes, which we analyze systematically. These relations
bring the properties of semi-standard Young tableaux
 of $SU(n-4)$ into the counting of basis amplitudes.

The representations we derive for superamplitudes take the form
\be
  \lab{cartoon}
  \ca_{n}^\text{N$^K$MHV} ~=~
  \sum_I A_I \, Z_I \, .
\ee
 The index $I$ enumerates the set of basis amplitudes $A_I$.
The $Z_I$ are SUSY and R-symmetry invariant Grassmann polynomials of degree $\cn(K+2)$.  They are constructed from two simple and familiar ingredients.  First each $Z_I$ contains a factor of
the well-known Grassmann delta-function,
 $\d^{(2\cn)}(\tilde Q_a)$,
which expresses the conservation of $\tilde Q_a$. It is annihilated by both $Q^a$ and $\tilde Q_a$.  The
 other ingredient is the first-order polynomial
\begin{equation}\label{introdan}
    \pdan_{ijk,a} \equiv [ij] \h_{ka} +[jk] \h_{ia} +[ki] \h_{ja}\,,
\end{equation}
which is annihilated by $Q^a$.
The $\eta_{ia}$  are the Grassmann
bookkeeping variables of superamplitudes, with  $a = 1,\ldots \cn$   the $SU(\cn)_R$ index, and
 with $i,j,k$ labeling three external lines of the $n$-point amplitude.
This polynomial
is the essential element of the well-known 3-point anti-MHV superamplitude.
Each $Z_I$ contains $\cn K$ factors of the polynomials $\pdan_{ijk,a}$.

Any $n$-point amplitude of the N$^K$MHV sector can be extracted from the superamplitude \reef{cartoon}. However, it is important to recognize that the sum over $I$
 in \reef{cartoon} does not span all possible amplitudes, but only a
 linearly independent subset, which we call  the \emph{algebraic basis} of amplitudes. In $\cn=4$ SYM theory,
the $\tilde{Q}_a$ Ward identities allow us to fix two states in the basis amplitudes to be negative-helicity gluons.\footnote{
Similarly, two states were fixed by SUSY to be gluons in \cite{ArkaniHamed:2008gz} in order to
establish the super-BCFW recursion relations~[\citen{NimaParis},\citen{Brandhuber:2008pf},\citen{ArkaniHamed:2008gz}] for tree amplitudes.
}
Likewise, $Q^a$ can be used to choose two other states to be positive-helicity gluons. We describe in detail how this solves the SUSY Ward identities. A priori, the remaining $n-4$  states are arbitrary particles of the theory: we write
\be
   \lab{cIs}
   A_I ~\longrightarrow~
   A_{n}(\,X_1\,\,X_2\,\dots\,X_{n-4}\,\,+\,+\,-\,-\,) \, ,
\ee
where $X_i$ denote $n-4$ states from the $\cn=4$ SYM multiplet.
The $X_i$ are restricted so that the amplitude belongs to the N$^K$MHV sector, as will be explained in the main text.

Amplitudes of the form $A_{n}(X_1\,X_2\dots X_{n-4}++--)$ are not all independent.
We will show that $SU(4)_R$ invariance of the superamplitude imposes multi-term linear relations among amplitudes whose external states are exactly of the same particle type but have different assignments of $SU(4)_R$ indices.
 An example is the
 following 4-term relation among NMHV amplitudes with gluinos $\l$ and scalars $s$:
\bea
  \nonumber
  0 &=&A_{6}( \l^{123} \l^{\bf 4} \l^{123} \l^{\bf 4} s^{1{\bf 3}} s^{2{\bf 4}} )
  ~+~ A_{6}( \l^{123} \l^{\bf 4} \l^{123} \l^{\bf 4} s^{1{\bf 4}} s^{2{\bf 3}} ) \\[1mm]
  &&
  ~+~ A_{6}( \l^{123} \l^{\bf 4} \l^{123} \l^{\bf 3} s^{1{\bf 4}} s^{2{\bf 4}} )
  ~+~ A_{6}( \l^{123} \l^{\bf 3} \l^{123} \l^{\bf 4} s^{1{\bf 4}} s^{2{\bf 4}} ) \, .
  \lab{6ptexI}
\eea
We call this a \emph{cyclic identity} because the four boldfaced $SU(4)_R$ indices are cyclically permuted.  Such identities also hold among
 amplitudes of the form $A_{n}(X_1\,X_2\dots X_{n-4}++--)$. For example \reef{6ptexI} becomes a relation among 10-point N$^3$MHV amplitudes by including the designated $++--$ gluons.
When considering all possible assignments of the $n-4$ states $X_i$ of \reef{cIs}, one should include in the algebraic basis only arrangements that cannot be related by $SU(4)_R$ symmetry.
Finding the linearly independent amplitudes can be formulated as a group theoretic problem and it has a neat solution.  Let us simply state the result for the N$^K$MHV sector: \emph{the number of amplitudes in the algebraic basis is the dimension of the irreducible representation of $SU(n-4)$ corresponding to a rectangular Young diagram with $K$ rows and
 $\cn$ columns!
The independent amplitudes are precisely labeled by the semi-standard tableaux of this Young diagram.} We will demonstrate this structure
 in more detail
for the NMHV and N$^2$MHV sectors in the summary below.

The number of amplitudes in the algebraic basis does not necessarily represent the minimal number of amplitudes one must compute in order to fully determine
a given $n$-point N$^K$MHV sector at any loop order. In $\cn=4$ SYM theory, the  single-trace color-ordered amplitudes have cyclic and reflection symmetries.
With the help of these symmetries, one amplitude can be related to another by a permutation of the external momenta\footnote{
At tree-level, there are additional functional relations between color-ordered amplitudes in gauge theory~\cite{Berends:1987me,Kleiss:1988ne,Bern:2008qj,BjerrumBohr:2009rd,Stieberger:2009hq}. We will not analyze the further reduction of the functional basis at tree-level due to these relations.}
For example, consider 6-point amplitudes of the NMHV sector in $\cn=4$ SYM. The algebraic basis can be chosen to be the 5 amplitudes
$A_6(  \pm \mp +--+)$, $A_6( \l^\pm \l^\mp +--+)$, and $A_6( s^{12} s^{34} +--+)$.
Cyclic and reflection symmetry relate the two gluon amplitudes as well as the two gluino amplitudes. Thus only 3 independent functions are needed to determine all $\approx 400$
 single-trace color-ordered amplitudes of the 6-point NMHV sector in $\cn=4$ SYM.

In $\cn = 8$ supergravity, the algebraic basis is defined as in $\cn = 4$ SYM, except that the basis amplitudes are now associated with the semi-standard tableaux of rectangular Young diagrams with $K$ rows and $8$ columns. However, an important difference is that gravity amplitudes are unordered, so all amplitudes with the same set of external particles
 and $SU(\cn)_R$ quantum numbers can be related by permutations of the momenta. For the $\cn = 8$ theory this defines
the \emph{functional basis},  and it is significantly smaller than the algebraic basis. For the NMHV sector with $n=6$ external particles, the algebraic basis consists of 9 amplitudes, but only the five listed in \reef{NMHVba} are functionally independent. For $n=7$, the reduction is from 45 amplitudes in the algebraic basis to just 10 in the functional basis. Beyond the NMHV level, the counting of functionally independent amplitudes becomes more difficult; we outline the procedure in Section \ref{sec8}.

We first encountered the problem of solving SUSY Ward identities for on-shell amplitudes in the 1977 work of Grisaru and Pendleton \cite{gris}. In addition to deriving the constraints on MHV amplitudes, the authors solved the SUSY Ward identities for the NMHV sector of $6$-point amplitudes of an $\cn=1$ supersymmetric theory.\footnote{The solutions of \cite{gris} were
rederived using spinor helicity methods in \cite{BEF}.} They found that six amplitudes were needed to determine all 60 NMHV amplitudes. Since the work \cite{gris}, there have been no similar systematic investigations.\footnote{MHV identities have been exploited repeatedly by many authors, but beyond the MHV level, we have only found partial results in the literature \cite{PTsusy,kunszt,Bidder:2005in}.} Given the recent interest in the maximally supersymmetric theories, $\cn=4$ SYM and $\cn=8$ supergravity, and the explicit amplitude calculations performed at tree and loop level, it is relevant to develop a clear and systematic understanding of
the consequences of supersymmetry and R-symmetry for amplitudes in these theories.

The superamplitude formulas can be adapted to open and closed superstring amplitudes with massless external states.
This follows from the results of Stieberger and Taylor \cite{ST2007}.
The SUSY Ward identities have been used in
explicit calculations of string amplitudes \cite{ST2007} and in
demonstrating the absence of certain higher order terms in the
$\alpha'$-expansion of closed string amplitudes
\cite{Stieberger2009,Obers2008}.

Higher loop calculations  in $\cn=4$ SYM  \cite{N4loop} and tests of the ultraviolet properties of $\cn=8$ supergravity \cite{Bern:2007hh,Bern:2008pv,Bern:2009kd} are currently done for MHV amplitudes.  Eventually calculations beyond the MHV level  may be needed, and it would be good to know how many independent amplitudes must be calculated. Superspace counterterms \cite{susyCTs} also provide useful information on this question.

In $\cn=4$ SYM, the BDS ansatz~\cite{Bern:2005iz,Anastasiou:2003kj} is believed to correctly reproduce planar 4- and 5-point MHV amplitudes to arbitrary loop order. However, the ansatz must be modified for MHV amplitudes with $n\geq 6$ external legs~\cite{Alday:2007he,Drummond:2007bm,N4loop,Anastasiou:2009kna}, and no similar ansatz at general N$^K$MHV level is known. We hope that our manifestly SUSY and R-invariant superamplitudes will help generalize the BDS conjecture to N$^K$MHV amplitudes.

Our results may also facilitate the evaluation of intermediate state sums needed in higher-loop calculations. Such sums are most efficiently performed using superamplitudes \cite{BEF,EFK1,EFK2, Bern:2009xq}. Perhaps some of the remarkable cancellations will naturally appear when the superamplitudes are written in a manifestly SUSY and R-invariant fashion.  We hope future work will shed light on these properties.

\subsection*{Summary of results}
N$^K$MHV superamplitudes take the general form~(\ref{cartoon}), with the index $I$ enumerating the semi-standard tableaux of the rectangular $SU(n-4)$ Young diagram with
 $K$ rows and
 $\cn$ columns.
These basis amplitudes are  multiplied by a manifestly SUSY and $SU(\cn)_R$-invariant $\eta$-polynomial. We now present the NMHV and N$^2$MHV superamplitudes of $\cn=4$ SYM  to give a  concrete illustration of this  structure.

At $n$-point NMHV level,  the amplitudes of the algebraic basis are characterized by four integers $1\!\leq\! i\!\leq\! j\!\leq\! k\!\leq\! l\!\leq\! n\!-\!4$\,, which correspond to the semi-standard tableaux
\raisebox{1pt}{
$\framebox[4mm][c]{\scriptsize $i$\phantom{j}\hspace{-1mm}}
  \framebox[4mm][c]{\scriptsize $j$\phantom{j}\hspace{-1mm}}
  \framebox[4mm][c]{\scriptsize $k$\phantom{j}\hspace{-1mm}}
  \framebox[4mm][c]{\scriptsize $l$\phantom{j}\hspace{-1mm}}$}
of a Young diagram with one row and four columns.
The basis amplitudes are  $A_n(\{i,j,k,l\}++--)$, with positive-helicity gluons on lines $n-3$ and $n-2$\,, negative-helicity gluons on lines $n-1$ and $n$. The notation $\{i,j,k,l\}$ dictates that $SU(4)_R$ index 1 is carried by line $i$,
 $SU(4)_R$ index 2 by line $j$, etc.  The particles on lines
that are not  included in a particular set  $\{i,j,k,l\}$ are  positive-helicity gluons. For example, $A_7(\{1,1,1,3\}++--) = A_7( \lambda^{123} + \lambda^{4} ++--)$.
The superamplitude  takes the form
\begin{equation}\label{introsup}
    \ca^{\rm NMHV}_n=\!\!\!\sum_{1\leq i\leq j\leq k\leq l\leq n-4}\!\!\!
    A_n(\{i,j,k,l\}++--) \,\,X_{(ijkl)}\,,
\end{equation}
where $X_{(ijkl)}$ is the total symmetrization\footnote{We define total symmetrization as the sum over inequivalent permutations, with no combinatorial factor. For example, $X_{(1112)}=X_{1112}+X_{1121}+X_{1211}+X_{2111}$\,.}  of the $\eta$-polynomial
\begin{equation}\label{introdefXs}
 X_{ijkl} ~\equiv~ \d^{(8)}\Bigl(\,\sum_{i=1}^n|i\>\eta_{ia}\Bigr)
 ~\frac{\,m_{i,n\dash3,n\dash2;1}\,\,m_{j,n\dash3,n\dash2;2}\,\,m_{k,n\dash3,n\dash2;3}\,\,m_{l,n\dash3,n\dash2;4}\,}{[n-3,n-2]^4\<n-1,n\>^4}\,.
\end{equation}
 Here, $\d^{(8)}$ is the Grassmann delta function and $m_{ijk,l}$ are the polynomials~(\ref{introdan}). We will elaborate on these functions in Section \ref{secsusyR}. The $X_{(ijkl)}$ are manifestly SUSY and R-symmetry invariant, and so is the superamplitude~(\ref{introsup}).

At   the N$^2$MHV level,  basis amplitudes are labeled by $SU(n-4)$ semi-standard Young tableaux with two rows and four  columns,
\begin{equation}
\begin{tabular}{|c|c|c|c|}
  \hline
  $\!\!i_1\!\!\!$ & $\!\!j_1\!\!\!$ & $\!\!k_1\!\!\!$ & $\!\!l_1\!\!$ \\
  \hline
  $\!\!i_2\!\!\!$ & $\!\!j_2\!\!\!$ & $\!\!k_2\!\!\!$ & $\!\!l_2\!\!$ \\
  \hline
\end{tabular}\,\,\,.
\end{equation}
Each row is non-decreasing ($i_A\!\leq\! j_A\! \leq\! k_A\!\leq\! l_A$) and each column is strictly increasing ($i_1\!<\!i_2$\,, etc.). From the hook rule \cite{hook} it follows that there are $(n\!-\!5)(n\!-\!4)^2(n\!-\!3)^2(n\!-\!2)^2(n\!-\!1)/(4!\,5!)$ semi-standard tableaux. Each tableau corresponds to an amplitude
$A_n\bigl(\,
\bigl\{
  {}^{i_1 j_1  k_1  l_1}_{i_2 j_2  k_2  l_2}
\bigr\}
\,++--\,\bigr)$
with the specified gluons on the last four lines and with $SU(4)_R$ index $1$ on lines $i_1$ and $i_2$, $SU(4)_R$ index $2$ on lines $j_1$ and $j_2$, etc.   For example,
\begin{equation}
   A_8( \lambda^{123} \lambda^{123} \lambda^{4} \lambda^{4}
  ++--)
  ~=~ A_8\bigl(
     \bigl\{
    {}^{1113}_{2224}
    \bigr\}
     ++--\bigr)~
~~~~\longleftrightarrow~~~~
   \begin{array}{l}
  \framebox[3.5mm][c]{\scriptsize 1}
  \framebox[3.5mm][c]{\scriptsize 1}
  \framebox[3.5mm][c]{\scriptsize 1}
  \framebox[3.5mm][c]{\scriptsize 3}\\[-2pt]
  \framebox[3.5mm][c]{\scriptsize 2}
  \framebox[3.5mm][c]{\scriptsize 2}
  \framebox[3.5mm][c]{\scriptsize 2}
  \framebox[3.5mm][c]{\scriptsize 4}
  \end{array}.
\end{equation}
The N$^2$MHV superamplitude can then be written in terms of  basis amplitudes as
\begin{equation}
   \ca^{\text{N$^2$MHV}}_n ~=~
   \frac{1}{16}
     \sum_{{}^\text{semi-standard}_\text{~~tableaux Y}}\!\!\! (-)^Y
     A_n\bigl(\,
     \bigl\{
    {}^{i_1 j_1  k_1  l_1}_{i_2 j_2  k_2  l_2}
    \bigr\}
     ++--\bigr)~
      Z^{i_1 j_1 k_1 l_1}_{i_ 2j_ 2k_2 l_2}
     \,,
\end{equation}
where the $Z$'s are
 manifestly SUSY- and R-symmetry invariant $\eta$-polynomials similar to the
 $X$'s in~(\ref{introdefXs}), but contain eight instead of four powers of $m_{ijk,a}$. The $Z$-polynomials and the sign factor $(-)^Y$ are defined  in Section~\ref{secN2MHVb}.

All basis amplitudes so far described carried gluons or gravitons  on four fixed external lines.
This choice was made when we solved the SUSY Ward identities,  but there are other possibilities.
For example, the Ward identity analysis
accommodates a basis in which any four specified lines carry the particle
content of an arbitrary $4$-point MHV amplitude. Thus, in $\cn=4$
SYM, one could pick all basis
 amplitudes to be of  the form
$A_{n}(\lambda^1\,X_1\,\lambda^{234}\,X_2\,s^{12}\,X_3\,s^{34}\,X_4\cdots X_{n-4})$\,. Even more general bases can be constructed. For example, one can choose
 a basis for $8$-point $\cn=4$ SYM amplitudes in which \emph{none} of the
 basis amplitudes contain gluons. Similarly, basis amplitudes without
gravitons are possible in supergravity up to $n=16$ external lines. Bases
without explicit gluons or gravitons may facilitate higher loop
perturbative calculations.\footnote{We thank Z. Bern for pointing this out.}
The
construction of more general bases is outlined in Appendix A. In the main
text we consider only basis
 amplitudes of the form
$A_{n}(X_1X_2\cdots X_{n-4}++--)$.  An advantage of this basis is that  the four fixed gluons or gravitons
are singlets of the R-symmetry, so this form is preserved under $SU(\cn)_R$  transformations.

\noindent {\sl Note added (September, 2010):}
The algebraic basis of the form $A_{n}(X_1X_2\cdots X_{n-4}++--)$ is particularly convenient to write down the superamplitude in closed form. Using a computer-based implementation of this superamplitude, however, one can choose any other set with the same number of linearly independent amplitudes. Linear independence, in this case, is best verified numerically. At the $6$-point NMHV level, for example, a suitable basis of $5$ linearly independent gauge theory amplitudes is the split-helicity amplitude $A_{6}(+++---)$ together with $4$ of its cyclic permutations. For $\cn=8$ supergravity, the graviton amplitude $M_{6}(+++---)$ together with $8$ permutations of its external lines  represents a suitable basis. It is striking that the functional basis at the $6$-point NMHV level (but not beyond!) can be reduced to a single all-gluon (or all-graviton) amplitude.

\subsection*{Organization of this paper}
The structure of the paper is as follows. In Section \ref{secsusyR}, we review the action of supersymmetry and R-symmetry on superamplitudes. In particular, we remind the reader why the MHV superamplitude is invariant under these symmetries.
Section \ref{secNMHV} is devoted to  NMHV superamplitudes.
We show how the solution to the SUSY Ward identities brings these superamplitudes to a manifestly supersymmetric form, and we present detailed examples and general formulas for superamplitudes in both $\cn=4$ SYM and $\cn=8$ supergravity.
The strategy for solving the SUSY Ward identities is very similar beyond the NMHV level. Now, however,
the role of R-symmetry becomes more central. We focus most of our discussion in Section \ref{secbeyond} on the N$^2$MHV sector of $\cn=4$ SYM theory, but results are given for the general  N$^K$MHV case. Section \ref{sec8} adapts the Yang-Mills results to $\cn=8$ supergravity. Appendix \ref{app18} briefly outlines how the NMHV superamplitude can be written in a completely general basis. We have collected some needed properties of Young tableaux in Appendix \ref{app17}.

\setcounter{equation}{0}
\section{Symmetries of superamplitudes}
\lab{secsusyR}

\subsection{Supersymmetry}
Particle states in the $\cn=4$ and $\cn=8$ theories are described by annihilation operators $A^{a_1 \cdots a_r}_i$, which carry the indices of a rank $0 \le r \le \cn$ fully antisymmetric representation of the global R-symmetry group $SU(\cn)_R$. The helicity of a particle  is then given by
$h = \cn/4-r/2$. The label $i$ denotes the momentum $p_i$.
 Thus we have 16 massless particles --- gluons, gluini, and scalars --- in the multiplet of $\cn=4$ SYM theory.
In  $\cn=8$ supergravity the
multiplet contains 256 states.

It is a consequence of R-symmetry that any N$^K$MHV amplitude contains a set of particles in which each index  $a=1,2,\ldots \cn$
appears $K+2$ times. For example, in $\cn=4$ SYM theory,
 $\big\< A_1 A^{34}_2 A^2_3 A^1_4 A^{1234}_5\big\>$ denotes a  5-point MHV amplitude in which the
particles are (in order) a positive-helicity gluon, a scalar, two positive-helicity gluinos, and a negative-helicity gluon.

Superamplitudes contain Grassmann variables $\h_{ia}$.  The $n$-point N$^K$MHV superamplitude $\ca^{\text{N$^K$MHV}}_n$ is a polynomial of degree $\cn(K+2)$ in the $\eta_{ia}$.
The supercharges, $Q^a = \e^\a Q_\a^a $ and $\tQ_a=\tilde \e_{\dot \a}\tilde Q_a^{\dot \a}$, act on a superamplitude by
multiplication or differentiation,
\be   \lab{tqq}
\tilde Q_a = \sum_{i=1}^n \< \eps\, i\>\, \eta_{ia}\,, \qquad\quad
Q^a = \sum_{i=1}^n [\eps \, i]\, \frac{\pa}{\pa\eta_{ia}} \, .
\ee

Every superamplitude with $n\ge 4$ external particles contains the factor (defined without $\<\eps|$)
\be \lab{deltq}
\d^{(2\cn)}(\tilde Q_a) ~\equiv~ \d^{(2\cn)}\Big(\sum_{i=1}^n |i\>\eta_{ia}\Big)
~=~ \frac{1}{2^{\cn}} \prod_{a=1}^{\cn}
\sum_{i,j=1}^n \<i\,j\>\, \h_{ia}\h_{ja}\,,
\ee
which expresses the conservation of $\tilde Q_a$. Indeed, it is clear that
$\tilde Q_a\, \d^{(2\cn)}\big(\tilde Q_a\big)  =0$.  Momentum conservation ensures that $Q^a\, \d^{(2\cn)}\big(\tilde Q_a\big) =0$.

The polynomials
\be \lab{pdef}
\pdan_{ijk,a} \equiv [ij] \h_{ka} +[jk] \h_{ia} +[ki] \h_{ja}\,
\ee
satisfy $Q^a \,\pdan_{ijk,b} =0$.
Since this relies only on the Schouten identity, it holds for any choice of three lines $i,j,k$, adjacent or non-adjacent, and is independent of
momentum conservation. Generally, $\tilde Q_a \pdan_{ijk,a} \ne 0$. However, if
 $p_i\!+\!p_j\!+\!p_k=0$\,, then $\tilde{Q}_a\, \pdan_{ijk,a} = 0$.

The Grassmann functions
 $\d^{(2\cn)}(\tilde Q_a)$
and $\pdan_{ijk,a}$ are familiar from MHV  and anti-MHV superamplitudes:
\be
  \lab{simple}
  \ca_n^\text{MHV}
  = A_{n}(++\dots+ - -)~ \frac{\d^{(2\cn)}(\tilde Q_a)}{\<n-1,n\>^{\cn}} \, ,
  \hspace{8mm}
  \ca_3^\text{anti-MHV}
  = A_{3}(-++) ~\prod_{a=1}^\cn \frac{\pdan_{123,a}}{[23]}\, .
\ee
Here, a pure gluon/graviton amplitudes appears as an overall factor because
 all MHV (anti-MHV) amplitudes are proportional to this amplitude.
This is a consequence of the SUSY Ward identities and therefore holds at  arbitrary loop order. It is also worth noting that choosing the MHV
``basis amplitude'' to be $A_{n}(++\dots+--)$ is a selection that makes lines $n-1$ and $n$
 special. This is compensated by the factor $1/\<n-1,n\>^\cn$. Beyond MHV level, we also select basis amplitudes with particular lines singled out. Similar compensator factors appear in our expressions.

 In the following sections we will see that SUSY Ward identities naturally lead us  to an expression for the N$^K$MHV superamplitude  as a sums of terms,
 each containing $\mathcal{N}\cdot K$ factors of the $\pdan_{ijk,a}$.  There is also an overall factor
of $\d^{(2\cn)}(\tilde Q_a)$. Such a polynomial is manifestly annihilated by all $Q^a$ and $\tilde Q_a$.
However, there are also important constraints from $SU(\mathcal{N})_R$ invariance which are discussed next.

\subsection{R-symmetry}
\lab{s:Rsym}

To establish $SU(\mathcal{N})_R$ invariance of a function of the $\eta_{ia}$-variables it is sufficient to impose invariance under $SU(2)_R$ transformations acting on any pair of the $SU(\mathcal{N})_R$ indices $1,\dots, \cn$.
To be specific, let us consider infinitesimal $SU(2)_R$ transformations in the $ab$-plane:
\be
  \lab{su2}
  \s_1:
  \Big\{
  \begin{array}{ccc}
  \delta_R \eta_{ia} &=& \theta \eta_{ib}\\
  \delta_R \eta_{ib} &=& \theta \eta_{ia}
  \end{array}
  \, ,
  \hspace{5mm}
  \s_2:
  \Big\{
  \begin{array}{ccc}
  \delta_R \eta_{ia} &=& -i \theta \eta_{ib}\\
  \delta_R \eta_{ib} &=& i \theta \eta_{ia}
  \end{array}
  \, ,
  \hspace{5mm}
  \s_3:
  \Big\{
  \begin{array}{ccc}
  \delta_R \eta_{ia} &=& \theta \eta_{ia}\\
  \delta_R \eta_{ib} &=& - \theta \eta_{ib }
  \end{array}
  \, .
\ee
Here $\theta$ is the infinitesimal transformation parameter.

As a warm-up to further applications, let us show that the MHV superamplitude is $SU(\cn)_R$-invariant. This simply requires that the $\d^{(2\cn)}$-function \reef{deltq} is invariant.
Since any monomial of the form $\eta_{i1}\,\eta_{j2}\cdots \eta_{l\cn}$ is invariant under a $\sigma_3$-transformation, so is the $\d^{(2\cn)}$-function. A $\s_1$-transformation in the 12-plane gives
\be
  \lab{dd}
  \delta_R\, \big( \d^{(2\cn)}(\tilde Q_a) \big)
  ~=~
  \frac{\theta}{2^{\cn-1}}
  \Big(
   \sum_{i,j=1}^n \< ij\> \, \eta_{i1} {\bf\eta_{j2}}~
   \sum_{k,l=1}^n \< kl\> \,{\bf \eta_{k2}  \eta_{ l2}} \Big)
   \Big(\prod_{a=3}^\mathcal{N}
  \sum_{k',l'=1}^n \< k'l'\> \eta_{k'a} \eta_{l'a}  \Big)~+ ~\dots
 ~~=~~ 0\, .~~~
\ee
Anticommutation of the (highlighted) Grassmann variables antisymmetrizes the sum over $j,k,l$ and $\< ij\>\< kl\>$ then vanishes by Schouten identity.
The ``$+\dots$" stands for independent terms from $\d_R$ acting   on $\eta_{k2}$ and $\eta_{l2}$. These terms can  be treated the same way. Invariance under  $\sigma_2$-transformations follows directly from $\sigma_{1,3}$-invariance.

Let us now consider the NMHV sector of the $\cn=4$ theory.
The superamplitude is manifestly $\tilde{Q}_a$-invariant when written as
\be
  \lab{Namp}
  \ca_n^\text{NMHV}~=~
  \delta^{(8)} \big( \tilde{Q}_a\big)\, P_4\, ,
  \hspace{1cm}
  P_4 ~= \sum_{i,j,k,l=1}^n q_{i j k l}\,
 \eta_{i 1} \, \eta_{j 2} \,  \eta_{k 3} \,  \eta_{l 4} \, .
\ee
Invariance under $\sigma_3$-transformations
 requires $P_4$ to be a linear combination of $\eta_{i1}\,\eta_{j2}\,\eta_{k3}\,\eta_{l4}$ monomials.
Consider the action of the $\sigma_1$-rotation
 in the $12$-plane:
\begin{equation}
    q_{ijkl}\,\d_R (\eta_{i1}\,\eta_{j2}\,\eta_{k3}\,\eta_{l4}) = \theta \, q_{ijkl}(\h_{i2}\h_{j2}+\h_{i1}\h_{j1})\eta_{k3}\,\eta_{l4}\,.
\end{equation}
This quantity must vanish; hence
$q_{ijkl}=q_{jikl}$. Similar arguments apply to any generator of $SU(4)_R$ and establish total symmetry of $q_{ijkl}$. We will use this property in Section \ref{secNMHV}. Beyond the NMHV level, the requirement of $SU(\cn)_R$ invariance imposes further relations. We describe this in detail  in Section \ref{secbeyond}.

\setcounter{equation}{0}
\section{NMHV superamplitude}
\label{secNMHV}

In this section we start with the $n$-point NMHV superamplitude of $\cn=4$ SYM in the general form \reef{Namp}. We impose the Ward identities to bring it to a manifestly supersymmetric and R-invariant form.
We identify the amplitudes in the algebraic basis. The reduction to the functional basis is given for single-trace amplitudes in Section \ref{secEx1}.
Section \ref{secEx2} presents the generalization to superamplitudes in $\cn=8$ supergravity.

\subsection{Solution to the NMHV SUSY Ward identities}
We start by rewriting the Grassmann $\delta^{(8)}$-function of \reef{Namp}.
As is well-known, it can be expressed as
\be
 \lab{first}
  \delta^{(8)} \Big( \sum_{i=1}^n |i\> \, \eta_{ia}\Big)
 ~=~
  \frac{1}{\<n-1,n\>^4}~
  \delta^{(4)} \Big( \sum_{i=1}^n \<n-1,i\> \, \eta_{ia}\Big)~
  \delta^{(4)} \Big( \sum_{j=1}^n \<n j\> \, \eta_{ja}\Big) \,,
\ee
using the Schouten identity.
The $\delta^{(4)}$-functions can be used to eliminate $\eta_{n-1,a}$ and $\eta_{na}$ from $P_4$; specifically
\bea\label{etaelim}
  \eta_{n-1, a} = -
  \sum_{i=1}^{n-2} \frac{\< n i \>}{\< n,n-1 \>} \, \eta_{i a}
  \, ,~~~~~
  \eta_{n a} =
  - \sum_{i=1}^{n-2} \frac{\< n-1, i \>}{\< n-1,n \>} \, \eta_{i a} \, .
\eea
Inserting this into the $P_4$ of \reef{Namp}, we find
\be
  \lab{step2}
 P_4 = \frac{1}{\<n-1,n\>^4} \sum_{i,j,k,l=1}^{n-2} c_{i j k l}\,
 \eta_{i 1} \, \eta_{j 2} \,  \eta_{k 3} \,  \eta_{l 4} \, .
\ee
The $c_{ijkl}$'s are linear combinations of the $q_{ijkl}$'s, but we will not need their detailed relationship. As above, R-symmetry requires the $c_{ijkl}$'s to be fully symmetric, so the number of needed inputs at this stage is $(n-2)(n-1)n(n+1)/4!$\,.

Next we impose the $Q^a$-Ward identities. We know that $Q^a$ annihilates
$\d^{(8)}\big( \tQ_a\big)$, so we are left to solve $Q^a\,P_4 = 0$. Consider the action of $Q^1$ on $P_4$:
\be
  0 ~=~ Q^1 P_4
     ~\propto~ \sum_{i,j,k,l=1}^{n-2} [\eps i ] \, c_{ijkl}\,
      \eta_{j2}\,\eta_{k3}\,\eta_{l4}
     ~=~
     \sum_{j,k,l=1}^{n-2} \Big[ \sum_{i=1}^{n-2} [\eps i ] \, c_{ijkl}  \Big]
      \eta_{j2}\,\eta_{k3}\,\eta_{l4}\, .
\ee
The quantity in square brackets must vanish for any triple $jkl$, so the $c_{ijkl}$ must satisfy
\be
  \lab{QWI}
  \sum_{i=1}^{n-2}\, [\eps i ] \, c_{ijkl}  ~=~ 0 \, .
\ee

 We now select two arbitrary (but fixed) lines $s$ and $t$ among the remaining lines $1,\ldots, n-2$.
We choose the SUSY spinor $|\e] \sim |t]$ and then  $|\e] \sim |s]$  and  use \reef{QWI} to
express the coefficients $c_{sjkl}$ and $c_{tjkl}$ in terms of $c_{ijkl}$ with
$i \neq s,t$:
\bea \lab{c1c2}
    c_{sjkl} = - \sum_{i\neq s,t}^{n-2} \frac{[t i ]}{[ts]} \, c_{ijkl}\, ,
    \hspace{5mm}
    c_{tjkl} = - \sum_{i\neq s,t}^{n-2} \frac{[s i ]}{[st]} \, c_{ijkl} \, .
\eea
The sums extend from $i=1$ to $i=n-2$, excluding lines $s$ and $t$.
Using supercharges $Q^a$, $a=2,3,4$\,, in the same way, we can write similar relations for
for $c_{iskl}, ~c_{itkl}$,  etc.

We use the relations~(\ref{c1c2}) to write $P_4$ in \reef{step2} as
 \bea
 \nonumber
 \<n-1,n\>^4 P_4 &=& \sum_{\,j,k,l = 1}^{n-2} \,\sum_{i\neq s,t}^{n-2} c_{ijkl}\,\,\eta_{i1} \eta_{j2}\,\eta_{k3}\,\eta_{l4} +\sum_{j,k,l=1}^{n-2}\big(
 c_{sjkl}\,\eta_{s1} + c_{tjkl}\,\eta_{t1}\big)\,\eta_{j2}\,\eta_{k3}\,\eta_{l4}\\
 &=& \frac{1}{[st]} \sum_{\,j,k,l = 1}^{n-2} \,\sum_{i\neq s,t}^{n-2} c_{ijkl}\, \pdan_{ist,1} \,\eta_{j2}\,\eta_{k3}\,\eta_{l4}\,,
 \lab{gdstuff}
 \eea
in which $\pdan_{ist,1}$ is the first-order polynomial
introduced in \reef{pdef}.   We repeat this process and use the analogues of \reef{c1c2} for
$c_{iskl}$ and $c_{itkl}$ to reexpress the
 sum over $j$ in \reef{gdstuff} in terms of $\pdan_{jst,2}$.
Repeating again for the $k$ and $l$ sums, we arrive
at a new form of the NMHV superamplitude that  is manifestly invariant under both $\tilde{Q}_a$
and $Q^a$ supersymmetry,
\be
  \lab{almostdone}
  \ca_n^\text{NMHV} = \sum_{i,j,k,l\neq s,t}^{n-2} c_{ijkl} \,X_{ijkl}\,,
 \ee
where we have introduced
\be \lab{xpoly}
  X_{ijkl} ~\equiv~ \d^{(8)}\big(\tQ_a\big)
  ~ \frac{m_{i s t,1}\,m_{j s t,2}\,m_{k s t,3}\,m_{l s t,4}}{[s t]^4\<n-1,n\>^4}\,.
\ee
The $\eta$-polynomial $X_{ijkl}$ of degree $12$ is a
remarkably simple function composed of the basic SUSY invariants
 that we introduced in Section~\ref{secsusyR}.

In the following, it is convenient to set $s=n-3$ and $t=n-2$.
Since the $c$-coefficients are fully symmetric we can symmetrize the $X$-polynomials and write
\be
   \lab{step35}
   \ca_n^\text{NMHV} ~=
     \sum_{1\le i\le j\le k\le l \le n-4}
     c_{ijkl}\,     X_{(ijkl)}   \, ,\hspace{1.3cm}
  X_{(ijkl)} \equiv \sum_{\mathcal{P}(i,j,k,l)} X_{ijkl}
\ee
The sum
 over permutations $\mathcal{P}(i,j,k,l)$ in the definition of $X_{(ijkl)}$ is over all \emph{distinct} arrangements of fixed indices $i,j,k,l$. For instance, we have $X_{(1112)}=X_{1112}+X_{1121}+X_{1211}+X_{2111}$. Also,
$X_{(1122)}$ contains the 6 distinct permutation of
 its indices, and $X_{(1123)}$ has 12 terms. The number of distinct permutations of a set with repeated entries is a multinomial coefficient \cite{stanley}.

Our final task is to identity the coefficients $c_{ijkl}$ as proportional to on-shell amplitudes of the basis.
Recall \cite{BEF} that component amplitudes are obtained by applying Grassmann derivatives to the superamplitude.
Consider amplitudes with negative-helicity gluons at positions $n-1$ and $n$. To extract such amplitudes from \reef{step35} we apply four $\eta_{n-1,a}$-derivatives and four $\eta_{na}$-derivatives to $\ca_n^\text{NMHV}$. These derivatives must hit the Grassmann $\d$-function and the result is simply a factor $\< n-1,n\>^4$, which cancels the same factor in the denominator of $X_{ijkl}$.
We must apply four more Grassmann derivatives
 $\frac{\pa}{\pa \eta_{i1}}\frac{\pa}{\pa \eta_{j2}}\frac{\pa}{\pa \eta_{k3}}\frac{\pa}{\pa \eta_{l4}}$ to $\ca_n$ in order to extract an NMHV amplitude. These derivatives hit the product of $\pdan_{i,n\dash3,n\dash2;a}$-polynomials and produce a factor of $[n\!-\!3,n\!-\!2]^4$ which cancels the
 remaining denominator factor of $X_{ijkl}$. As a result, the 12 $\eta$-derivatives just leave us with the coefficient $c_{ijkl}$. When
$\!1\!\le\! i\! \le\! j\! \le\! k\! \le\! l \!\le\! n\!-\!4$\,,
we have therefore identified $c_{ijkl}$ as the amplitude
\be \lab{ampl}
   c_{ijkl}  ~=~
   A_n \big( \{i,j,k,l\}++ -- \big)
   ~\equiv~
   \big\<  \cdots A_i^1 \cdots
     A_j^2 \cdots A_k^3 \cdots A_l^4 \cdots
   A_{n-1}^{1234} A_{n}^{1234}
   \big\>  \, .
\ee
Let us clarify the notation: $A_n \big(  \{i,j,k,l\} ++--\big)$ means that line $i$ carries the $SU(4)_R$ index 1, line $j$ carries index 2 etc.
If $i=j$, this means that
the line carries both indices 1 and 2, and the notation $A_i^1 A_i^2$ should then be understood as $A_{i}^{12}$. Furthermore the dots indicate positive-helicity gluons in the unspecified positions, specifically on line numbers
$1,2,..,i-1;$~$i+1,..,j-1;$~$j+1,..,k-1;$~$k+1,..,l-1$; and $l+1,..,n-2$.   In particular, there are positive-helicity gluons
on lines $n-3$ and $n-2$ in all amplitudes of this basis.

With this identification of the $c_{ijkl}$ coefficients we can now write
\be
   \lab{step4}
   \ca_n^\text{NMHV} ~=
     \sum_{1\le i\le j\le k\le l \le n-4}
      A_n \big(   \{i,j,k,l\}++ --\big)
     ~X_{(ijkl)}   \, ,
\ee
This is our final result for the NMHV superamplitude
 in $\cn=4$ SYM.
Any desired amplitude can be obtained by applying the 12th-order Grassmann derivative that corresponds to
its external states.
Note that the use of the $Q^a$ Ward identities has reduced the counting of independent  basis amplitudes
to \mbox{$(n-4)(n-3)(n-2)(n-1)/4!$}\,. This number is also the dimension of the fully symmetric 4-box irreducible representation of $SU(n-4)$.
 The significance of this will become clear when we discuss the N$^K$MHV sector.

The  representation \reef{step4} contains a sum over basis amplitudes which are \emph{algebraically} independent under the symmetries we have imposed.  However we have not yet exploited the dihedral symmetry of color-ordered amplitudes in $\cn=4$ SYM.   This imposes \emph{functional} relations among amplitudes,  specifically there are amplitude
relations involving reordering of  particle momenta.  Analogous relations appear in $\cn=8$ supergravity because there is no color-ordering in gravity. These features are discussed in the next two subsections.


\subsection{Single-trace amplitudes of $\cn=4$ SYM and the functional basis}
\lab{secEx1}

Let us write the 6-point superamplitude in the algebraic basis described above:
\bea
  \nonumber
   {\ca}_6^\text{NMHV} &=&
   \big\< -+++--\big\> \,\, X_{1111}
   ~+~ \big\<  \lambda^{123} \lambda^{4}++ --\big\>\, X_{(1112)}
   ~+~\big\< s^{12}\, s^{34} \,++--\big\>\,
   X_{(1122)}
   \\[1mm]
   &&~~~~~~
   +~  \big\<  \lambda^{1} \lambda^{234} ++--\big\> \,X_{(1222)}
   ~+~ \big\< +-++--\big\> \, X_{2222}
    \, .
    \lab{nmhv6}
\eea
Here, we use a shorthand notation where $+$ and $-$ denote gluons $A$ and $A^{1234}$, respectively, $\lambda$ denotes a gluinos ($A^a$ or $A^{abc}$)  with the indicated $SU(4)_R$ indices, and $s^{ab}$ denotes the scalar $A^{ab}$.

In \reef{nmhv6} five basis amplitudes were needed to determine the superamplitude. Amplitudes with a single color-trace structure have   cyclic and reflection symmetries, which make further reduction possible. To exploit this dihedral symmetry, we choose an algebraic basis in which states 1 and n are fixed to be negative-helicity gluons and states $2$ and $n-1$ are positive-helicity gluons:
\bea
  \nonumber
  &&\HEW^{(1)} = \< -_1 +_2-_3 +_4+_5-_6\> \, ,~~~
  \HEW^{(2)} = \< -_1 +_2 \lambda^{123}_3 \,\lambda^{4}_4 +_5-_6\> \,  ,~~~
  \HEW^{(3)} = \< -_1 +_2 s^{12}_3\, s^{34}_4+_5-_6\> \,  ,~~~\\[2mm]
  &&
  \HEW^{(4)} = \< -_1 +_2 \lambda^{1}_3 \,\lambda^{234}_4 +_5-_6\>\,  ,~~~
  \HEW^{(5)} = \< -_1 +_2 +_3 -_4+_5-_6\> \, .
\eea
(The subscript denotes the momentum label.)
These five amplitudes are not \emph{functionally}  independent:  $\HEW^{(5)}$ is related to $\HEW^{(1)}$,  and $\HEW^{(4)}$ to $\HEW^{(2)}$, by reflection, viz.
\bea
  \hspace{-5mm}
  && \HEW^{(5)}
  =  \< -_1 +_2 +_3 -_4+_5-_6\>
  = \< -_6 +_5 -_4 +_3+_2-_1\>
  = \mathcal{R} \,\HEW^{(1)}\, , \\[2mm]
\nonumber
  \hspace{-5mm}&& \HEW^{(4)}
  = \< -_1 +_2 \lambda^{1}_3 \,\lambda^{234}_4 +_5-_6\>
  = -\< -_1 +_2 \lambda^{4}_3 \,\,\lambda^{123}_4 +_5-_6\>
  = \< -_6 +_5 \lambda^{123}_4 \,\lambda^{4}_3 +_2-_1\>
  =  \mathcal{R} \, \HEW^{(2)} \, .
\eea
We have introduced the ``reversal operator'' $\mathcal{R}$, which leaves the states invariant but reverses the order of momenta.

Thus, at any loop order in the single-trace sector, one needs to determine only the three NMHV amplitudes $\HEW^{(1,2,3)}$ in order to know all  6-point NMHV amplitudes. We use the term \emph{functional basis} for a basis of amplitudes reduced using dihedral symmetry. In the functional basis, the 6-point NMHV superamplitude can be written
\be
   \ca_6^\text{NMHV} ~=~
   \HEW^{(1)} ~X_{3333}
   ~+~ \HEW^{(2)} ~X_{(3334)}
   ~+~ \HEW^{(3)} ~X_{(3344)}
   ~+~  (\mathcal{R} \, \HEW^{(2)}) ~X_{(3444)}
   ~+~  (\mathcal{R}\, \HEW^{(1)})~ X_{4444}
    \, .
\ee
In this formula, $X_{(ijkl)}$ are the symmetrized sums of $X_{ijkl} = [25]^{-4} \<16\>^{-4}\;\pdan_{i25,1}\; \pdan_{j25,2}\; \pdan_{k25,3}\; \pdan_{l25,4}$.

\noindent {\bf Functional basis for all $n$:}\\
For single-trace amplitudes, it is convenient to choose an algebraic basis of the type  $A_n(-+\,\cdot\,\,\cdot\,\,\cdot\, +-)$ to determine the functional basis.
For these amplitudes, the analysis of dihedral symmetry can actually be carried out for general $n$. In fact, any algebraic basis amplitude $A_n(-+\,\cdot\,\,\cdot\,\,\cdot\, +-)$ that is not invariant under reflection, ${\cal R}A_n\neq A_n$\,, is functionally related to the basis amplitude ${\cal R}A_n$ and we only need to keep one of these
two amplitudes to form a functional basis. Reflection-symmetric basis amplitudes, however, are not related to other basis amplitudes through the dihedral symmetry and are thus all part of the functional basis.
There are $(n^2\!-\!4n\!+\!6)(n\!-\!4)(n\!-\!2)/48$ functional basis amplitudes for even $n$\,, and $(n^2\!-\!6n\!+\!11)(n\!-\!3)(n\!-\!1)/48$ for odd $n$.

\subsection{NMHV amplitudes in $\cn=8$ supergravity}
\lab{secEx2}

The identification of an algebraic basis
 in supergravity proceeds as in gauge theory and leads
to a representation of  NMHV
superamplitudes analogous to \reef{step35},
namely
\be \lab{repgrav}
 \cm_n^\text{NMHV}\,=\!\!\!
  \sum_{1\le i \le j \le \dots \le v \le n\dash4}\!\!\!
  c_{ijklpquv}~
  X_{(ijklpquv)},
\ee
with symmetrized versions of the $Q^a$- and $\tQ_a$-invariant polynomial
\be \lab{xpoly8}
 X_{ijklpquv} ~=~ \d^{(16)}(\tQ_a)~
 \frac{\,\pdan_{i,n\dash 3,n\dash 2;1}\,\,\pdan_{j,n\dash 3,n\dash2;2}\,\cdots\,\pdan_{v,n\dash 3,n\dash 2;8}\,}{[n-3,n- 2]^8\<n-1,n\>^8 }\,\,.
\ee
 As in $\cn=4$ SYM, we can identify each
 coefficient $c_{ijklpquv}$ with an amplitude:
 \begin{equation}
    c_{ijklpquv}  ~=~
   M_n \big( \{i,j,k,l,p,q,u,v\}++ -- \big)
   ~\equiv~
   \big\<  \cdots A_i^1 \cdots
     A_j^2 \cdots\cdots  A_v^8 \cdots
   A_{n-1}^{12345678} A_{n}^{12345678}
   \big\>  \, .
 \end{equation}
The notation $\{i,j,k,l,p,q,u,v\}$  indicates that line
 $i$ carries $SU(8)_R$ index 1, while line $j$ carries $SU(8)_R$ index 2, etc.   There may be equalities such as $i = j$\,, which means that the line in question carries both $SU(8)_R$ indices $1$ and $2$.

In gravity, as opposed to gauge theory, there is no ordering of the external states. Therefore amplitudes with the same external particles
 and the same $SU(8)_R$ charges are all related by momentum relabeling. For example,
\be
  c_{22222222}~=~ M_6(+_1 -_2 +_3 +_4 -_5 -_6)
  = M_6(-_2 +_1 +_3 +_4 -_5 -_6)
  = \big(c_{11111111} ~~\text{with}~~p_1 \lra p_2\big)\,.
\ee
Thus the basis amplitudes are functionally independent only if the sets of external particles on lines 1 to $n-4$ are distinct.
Since there are a total of eight $SU(8)_R$ indices $1,2,\dots,8$  distributed on these $n-4$ states, the number of independent amplitudes is equal to the number of partitions of 8 into $n-4$ non-negative integers.

For example, for $n=6$ we have the partitions $[8,0]$\,, $[7,1]$\,, $[6,2]$\,, $[5,3]$ and $[4,4]$
corresponding to five amplitudes in the functional basis.
The 6-point superamplitude is then
\bea
  \nonumber
   {\cal M}_6^\text{NMHV} &=&~
   \Big\{~~
   \big\<    -\, + \,+ + - -\big\>\,\, X_{\,11111111\,}
   \,~+~ \big\<   \psi^{-}\psi^{+}  + + - -\big\>\,
   X_{(11111112)}
   \\[1mm]  \nonumber
   &&~\,
   +\big\<  v^-v^++ + - -\big\>\,
   X_{(11111122)}
   ~+~\big\< \chi^{-}\chi^{+} + + - -\big\>\,
   X_{(11111222)} \\[1mm]
   &&~\,
   +\frac{1}{2}\big\< \, \phi^{1234} \phi^{5678} ++  - -\big\>\,
   X_{(11112222)} \Big\}
   ~~+~~ \text{(1 $\leftrightarrow$ 2)}\, .
   \lab{SGNMHV}
\eea
Particle types are indicated by $\psi^+ = A^{1}$, $\psi^- = A^{2345678}$ etc, in hopefully self-explanatory notation.
The ``$\,+~(1 \lra 2)$'' exchanges momentum labels 1 and 2 in the $X$-polynomials as well as in the basis amplitudes. The exchange does not introduce new basis functions, it only relabels momenta in the basis amplitudes written explicitly in \reef{SGNMHV}.

For $n$-point amplitudes the count of partitions of 8, and thus of the number of functionally independent basis amplitudes, is

\begin{tabular}{rcccccccccccccccc}
  $n~~= $ & 5 & 6 & 7 & 8 & 9 & 10 & 11 & $\ge$ 12 \\[1mm]
  $\text{basis count}~
  \,= $  & 1 & 5 & 10 & 15 & 18 & 20 & 21 & ~~~22
\end{tabular}

\vspace{1mm}
\noindent
The entry in the second line is the number of $n$-point amplitudes one needs to compute in order to fully determine the $n$-point NMHV superamplitude.

The saturation at $n=12$ occurs because the longest partition of $n-4=8$ is reached, namely the partition $[1,1,1,1,1,1,1,1]$. This partition corresponds to a basis amplitude with 8 gravitinos, two positive-helicity gravitons and two negative-helicity gravitons.
 For $n>12$, one only adds further positive-helicity gluons to each partition. This does not change the count of independent amplitudes.

\vspace{2mm}
\noindent {\bf Examples} \\
Let us illustrate the solution to the Ward identities in a few explicit examples.  Consider first the amplitude with 2 sets of 3 identical gravitinos, $A_6(\psi^{1234567}\,  \psi^8\,  \psi^8\,  \psi^8 \, \psi^{1234567} \, \psi^{1234567})$. Applying the corresponding Grassmann derivatives \cite{BEF} to the superamplitude \reef{SGNMHV}, we find
\bea
 \nonumber
 &&\hspace{-2cm}
 M_6(\psi^{1234567}\, \psi^8 \, \psi^8 \, \psi^8\,  \psi^{1234567} \psi^{1234567})\\[1mm]
 &&=
 \frac{1}{[34]\<56\>} \,
 \Big\{
  ~ \< 2|3+4|1 ] ~\big\<    -\, + \,+ + - -\big\>
   + s_{234}  ~\big\<   \psi^{-}\psi^{+}  + + - -\big\>
 \Big\} \, ,
 \lab{ex1}
\eea
where $s_{234} = - (p_2 +p_3+p_4)^2$.
This particular $\cn=8$ amplitude agrees with the 6-gravitino amplitude
$\big\< \psi^{-}\psi^{+} \psi^{+} \psi^{+} \psi^{-} \psi^{-}\big\>$ in the
truncation of the $\cn=8$ theory to $\cn = 1$ supergravity. In fact the relation \reef{ex1} is a special case of the ``old'' solution to the $\cn=1$ SUSY Ward identities \cite{gris,BEF}.

An example which does not reduce to $\cn=1$ supergravity is obtained by interchanging
 the $SU(8)_R$ indices 7 and 8 on states 1 and 2 in the 6-gravitino amplitude. The result is another 6-gravitino amplitude whose expression in terms of basis amplitudes is found to be
\bea
 \nonumber
 &&\hspace{-2cm}
 M_6(\psi^{1234568}\, \psi^7 \, \psi^8 \, \psi^8\,  \psi^{1234567} \psi^{1234567})\\[1mm]
 &&=
 -\frac{1}{[34]\<56\>} \,
 \Big\{
  ~  s_{134}  ~\big\<   \psi^{-}\psi^{+}  + + - -\big\>
   +  \< 1|3+4|2 ] ~\big\<    v^-\, v^+ \,+ + - -\big\>
 \Big\} \, .
 \lab{ex2}
\eea
This example could be interpreted as the solution to the SUSY Ward identities in $\cn=2$ supergravity.

Our final example contains two distinct scalars and four gravitinos:
\bea
 \lab{ex3}
 &&\hspace{-1cm}
 M_6(\phi^{1238}\, \phi^{4568} \, \psi^7 \, \psi^8\,  \psi^{1234567} \psi^{1234567})\\[1mm]
 &&=
 \frac{\< 2|1+4|3 ]}{[34]^2\<56\>} \,
  \Big\{
  ~   [14]    ~\big\<   \chi^{-}\chi^{+}  + + - -\big\>
     + [24]   ~\big\<   \phi^{1234}\phi^{5678}  + + - -\big\>
 \Big\}  ~-~(1 \lra 2) \, .
 \nonumber
\eea
We have checked the solutions \reef{ex1}, \reef{ex2}, and \reef{ex3} numerically at tree level using the MHV vertex expansion, which is valid for the specific $\cn=8$ amplitudes considered here. Of course the relations \reef{ex1}, \reef{ex2}, and \reef{ex3} hold in general, at arbitrary loop order.

\setcounter{equation}{0}
\section{Beyond NMHV}\label{secbeyond}
We are now ready to venture beyond the NMHV level. The first part of the analysis, in Section \ref{secNKMHV}, follows the previous strategy for solving the SUSY constraints. It leads to a representation for N$^K$MHV superamplitudes in $\cn=4$
SYM that is similar to \reef{almostdone}. This representation satisfies
the $\tQ_a$ and $Q^a$ Ward identities, but it is not $SU(4)_R$-invariant
when $K >1$. Indeed, $SU(4)_R$ symmetry requires that the $c$-coefficients, and thus the basis amplitudes,
of N$^K$MHV superamplitudes for $K \ge 2$ satisfy $(K+1)$-term cyclic identities.
The cyclic identities are derived in
Section \ref{seccyclic}. This leads to the hypothesis and proof in Section \ref{secNKMHVa} that the algebraic basis for N$^K$MHV superamplitudes
corresponds to semi-standard tableaux of
 rectangular $K$-by-$\cn$ Young diagrams. Explicit
representations for superamplitudes are given in Section \ref{secN2MHVb}. The
analysis is extended to $\cn=8$ supergravity in Section \ref{sec8}.

\subsection{Solution to the N$^K$MHV SUSY Ward identities}
\lab{secNKMHV}
The $n$-point N$^K$MHV superamplitude of ${\cal N}=4$ SYM is a polynomial of order $4(K+2)$ in $\eta_{ia}$'s.
Each $SU(4)_R$
index $a=1,2,3,4$ must appear $K+2$ times in every term of the polynomial.
We can write it as
\be\label{NkMHVpoly}
  \ca^{\text{N$^K$MHV}}_n
  \,=\, \d^{(8)} \big( \tilde{Q}_a \big)\,P_{4K}\,,
\ee
where $P_{4K}$ is a degree $4K$ Grassmann polynomial
\begin{equation}\label{P4K}
    P_{4K}=\frac{1}{\< n-1,n \>^4}
    \frac{1}{K!^4}\!\!\sum_{i_A,j_A,k_A,l_A=1}^{n-2} \!\!c_{[i_A,j_A,k_A,l_A]}\, \prod_{B=1}^{K}\eta_{i_B,1}\eta_{j_B,2}\eta_{k_B,3}\eta_{l_B,4}\,.
\end{equation}
Here we have already used the Grassmann $\d$-function to eliminate $\eta_{n-1,a}$ and $\eta_{na}$ from $P_{4K}$, just as we did in the NMHV case.

It is useful to think of the $4K$ indices $[i_A,j_A,k_A,l_A]$ as an array of $K$ rows and $4$ columns:
\begin{equation}
\lab{array}
[i_A,j_A,k_A,l_A]
~=~~
    \begin{tabular}{|c|c|c|c|}
       \hline
       $\!i_1\!$ & $\!j_1\!$ & $\!k_1\!$ & $\!l_1\!$ \\
       $\vdots$ & $\vdots$ & $\vdots$ & $\vdots$ \\
       $\!i_K\!$ & $\!j_K\!$ & $\!k_K\!$ & $\!l_K\!$ \\
       \hline
     \end{tabular}~\,.
\end{equation}
The first column $i_A,   A=1,2,\dots K$ lists the particles that carry the $SU(4)_R$ index $a=1$. The column $j_A,   A=1,2,\dots K$ lists the particles that carry index $a=2$, etc.
The coefficients $c_{[i_A,j_A,k_A,l_A]}$ enjoy the following properties:
\begin{itemize}
  \item \emph{Each column is fully antisymmetric.}\\
Each monomial term in $P_{4K}$
is antisymmetric in the $K$ Grassmann variables $\eta_{i_A,1}$, and hence the $c$-coefficients can be assumed to be fully antisymmetric under exchange of any indices of the first column $i_A$. Similarly, each of the other columns $j_A$, $k_A$, $l_A$ are fully antisymmetric.

 \item \emph{Symmetry under exchange of full columns.}\\
Consider the finite $SU(2)_R \subset SU(4)_R$ group transformation, $\eta_{i1} \to \eta_{i2}$ and  $\eta_{i2} \to -\eta_{i1}$ for all $i=1,\dots, n$. This must leave the superamplitude invariant, hence $c_{[i_A,j_A,k_A,l_A]} = c_{[j_A,i_A,k_A,l_A]}$. This must hold for any $SU(2)_R \subset SU(4)_R$.
Therefore, the coefficients $c_{[i_A,j_A,k_A,l_A]}$ are invariant under any exchange of full columns of indices.

\item \emph{Cyclic identities.}\\
Invariance under infinitesimal $SU(4)_R$ transformations requires the coefficients to satisfy certain $(K+1)$-term cyclic identities.
We discuss these in Section \ref{seccyclic}.

\end{itemize}

Now we analyze the $Q^a = \sum_{i=1}^n [\eps\, i] \, \frac{\pa}{\pa \eta_{ia}}$ Ward identities. When $a=1$, the derivatives in $Q^1$ hit the $K$ $\eta_{i_A1}$'s in $P_{4K}$.
Since the indices in the first column $i_A$ of $c_{[i_A,j_A,k_A,l_A]}$ are antisymmetric and summation indices can be relabeled, each of the $K$ terms gives the same result. We can then write $Q^1 P_{4K}=0$ as
\begin{equation}
  \lab{Qk}
    \sum_{j_A,k_A,l_A=1}^{n-2}\,\sum_{i_2,\ldots,i_K=1}^{n-2}
    \Bigl\{\sum_{i_1=1}^{n-2} [\epsilon i_1]\,c_{[\{i_1,i_2,\ldots,i_K\},j_A,k_A,l_A]}\Bigr\}
    \eta_{i_2,1}\cdots \eta_{i_K,1}\prod_{B=1}^{K}\eta_{j_B,2}\eta_{k_B,3}\eta_{l_B,4}~~=~~0\,.~~~
\end{equation}
Note that the left hand side contains a distinct $\eta$-monomial for each choice of $i_2,\ldots,i_K,j_A,k_A,l_A$. To satisfy the $Q^a$ Ward identity, the coefficient of {\em each monomial} must vanish separately. It follows that the expression in braces  in \reef{Qk}
has to vanish for any  $i_2,\ldots,i_K,j_A,k_A,l_A$.
As in the NMHV sector, we use  these constraints to express any $c$ with
 $i_A=n-3$ or $i_A=n-2$ for some $A$ in terms of $c$'s with $1\le i_A \le n-4$ for all $A$.
A new feature is that the set $i_A$ can
 initially contain both indices $(n-3)$ and $(n-2)$, but
 these can be eliminated one after the other.  We can then write the superamplitude in terms of the $Q^a$-invariant  $\pdan_{ijk,a}$-polynomials:
\begin{equation}
  \lab{NKMHV1}
   \ca^{\text{N$^K$MHV}}_n ~=~
   \frac{1}{K!^4}
     \!\!\!\sum_{i_A,j_A, k_A,l_A=1}^{n-4}\!\!\!
     \,c_{[i_A,j_A,k_A,l_A]}\,
      X_{[i_A,j_A,k_A,l_A]}
     \, ,
\end{equation}
where
\be
  \lab{genX1}
  X_{[i_A,j_A,k_A,l_A]}\equiv
  \frac{\delta^{(8)}\big( \tilde{Q}_a \big)}{\<n-1,n\>^4}
       \prod_{B=1}^{K}
     \frac{\,\pdan_{i_B,n\dash3,n\dash2;1}\; \pdan_{j_B,n\dash3,n\dash2;2}\; \pdan_{k_B,n\dash3,n\dash2;3}\; \pdan_{l_B,n\dash3,n\dash2;4}\,}{[n-3,n-2]^4} \, \, .
\ee
Let us now describe the amplitudes corresponding to the $c$-coefficients in \reef{NKMHV1}.

One might now be tempted to conclude that the basis of independent coefficients are those $c_{[i_A,j_A,k_A,l_A]}$'s  with
(1) ordered columns $i_1 < i_2 < \dots < i_K$ etc, and
(2) $i_1 \le j_1 \le k_1 \le l_1$ (by set exchange symmetry), with (3) all indices in the range 1 to $n-4$.
However, the cyclic identities mentioned above impose further constraints among these coefficients, which we illustrate in detail in Section~\ref{seccyclic} for the N$^2$MHV superamplitude and then generalize to $K\ge2$. Before moving on to the analysis of cyclic identities, we devote the following section to the precise relationship between $c$-coefficients and amplitudes.

\subsection{Relationship between $c_{[i_A,j_A,k_A,l_A]}$-coefficients and amplitudes}
\lab{secAc}
An amplitude is projected out from
 a superamplitude by applying the Grassmann derivatives $\partial_{i}^{ab\dots} \equiv
\frac{\pa}{\pa \eta_{ia}} \frac{\pa}{\pa \eta_{ib}} \dots$ that correspond to its external particles \cite{BEF}.
 For example, $\partial_{i}^{1234}$ corresponds to a negative-helicity gluon on line $i$. Projecting out an amplitude of the form $A_n(\cdots ++--)$ from the superamplitude \reef{NKMHV1}, we find that all angle brackets $\<n\!-\!1,1\>$ and square brackets $[n\!-\!3,n\!-\!2]$ cancel, just as in the NMHV case (see discussion above~(\ref{ampl})\,). We thus find that each $c$-coefficient gives precisely one amplitude of the form $A_n(\cdots ++--)$, up to a possible sign. The amplitude corresponding to a $c$-coefficient thus carries the states $++--$ on the last four lines; the particles on the remaining lines are determined as follows.
The indices of $c_{[i_A,j_A,k_A,l_A]}$ indicate how $K$ sets of $SU(4)_R$
 indices are distributed among the $n-4$ first states: states $i_1,\dots,i_K$ carry $SU(4)_R$ index 1, states $j_1,\dots,j_K$ carry $SU(4)_R$ index 2, etc.
Introducing the shorthand $c^{i_1j_1k_1l_1}_{i_2j_2k_2l_2}$ for $c$-coefficients at N$^2$MHV level we find,
for example,\footnote{
To extract this amplitude, we apply $\pa_1^{123} \pa_2^{123} \pa_4^4 \pa_5^4 \pa_8^{1234} \pa_9^{1234}$ to the superamplitude \reef{NKMHV1}. The last two derivatives hit only the Grassmann delta function and produce a factor of $\<89\>^4$. The remaining derivatives hit the $m_{i67,a}$-polynomials in $c^{1114}_{2225}\, X^{1114}_{2225}$ and produce $[67]^8$. Since $\<89\>^4 [67]^8$ cancels the equivalent denominator factor in \reef{genX1}, we are left with just the coefficient
$c^{1114}_{2225}$ and a minus sign produced by the Grassmann differentiations.}
$c^{1114}_{2225} =
  -\, A_{9}( \lambda^{123} \lambda^{123} + \lambda^{4} \,\lambda^{4} ++ --  )$.

For $K=2$ we write
\be
  \lab{ctoA}
  c^{i_1 j_1  k_1  l_1}_{i_2 j_2  k_2  l_2} =
  (-)^Y ~
  A_n\bigl(\,\bigl\{{}^{i_1 j_1  k_1  l_1}_{i_2 j_2  k_2  l_2} \bigr\}++--\bigr)\,,
\ee
where $(-)^Y$ is the sign factor $\pm1$ that compensates for a possible difference in sign between the $c$-coefficient and the amplitude.\footnote{
We
have $(-)^Y = \mathrm{sign}{(\s)}$, where $\s$ is the permutation that
places $\{i_1,\ldots,i_K,j_1,\ldots,j_K,k_1,\ldots,
k_K,l_1,\ldots,l_K\}$ in numerical order. If two or more indices
coincide, the permutation
is unique only up to a residual permutation of identical indices. In this case we demand that the permutation preserves the initial
order of identical indices.
 For example, for $Y=\bigl\{{}^{1113}_{2224}\bigr\}$ we have $(-)^Y=-1$
because an odd permutation is needed to turn $\{1,2,1',2',1'',2'',3,4\}$
into $\{1,1',1'',2,2',2'',3,4\}$\,.}

\subsection{Cyclic identities}
\lab{seccyclic}

It is not difficult  to derive the cyclic identities advertised earlier. We first treat  N$^2$MHV superamplitudes in the early form \reef{P4K} where $P_{4K}=P_8$ is an 8th order polynomial.
Consider the infinitesimal $\sigma_1$ transformation  of \reef{su2} acting in the 34 plane.   R-symmetry requires the variation $\delta_R \,  P_8$ to vanish, i.e.
\bea
  \nonumber
  0~=~\delta_R \,  P_8 &\propto&
  \sum_{i_A,j_A,k_A,l_A=1}^{n-2} \!\!
  c^{i_1 j_1 k_1 l_1}_{i_2 j_2 k_2 l_2}~\,
  \eta_{i_1,1}\, \eta_{i_2,1}\,
  \eta_{j_1,2}\, \eta_{j_2,2}\,
  ~\delta_R \big(
  \eta_{k_1,3}\, \eta_{k_2,3}\,
  \eta_{l_1,4}\, \eta_{l_2,4} \big)\\
  &=&
  2 \, \theta \sum_{i_A,j_A,k_A,l_A=1}^{n-2} \!\!
   c^{i_1 j_1 k_1 {\bf l_1}}_{i_2 j_2 {\bf k_2 l_2}}~\,
  \eta_{i_1,1}\, \eta_{i_2,1}\,
  \eta_{j_1,2}\, \eta_{j_2,2}\,
  \eta_{k_1,3}\,
  \eta_{{\bf k_2},4}\, \eta_{{\bf l_1},4}\, \eta_{{\bf l_2},4}
  + \dots \, .
\eea
(The $+\dots$ indicate terms in which the $\eta_{i4}$ are transformed;
they vanish independently.) After antisymmetrizing the indices $k_2, l_1, l_2$\,, we find the  \emph{cyclic identity}
\be
\lab{cyc1}
   c^{i_1 j_1 k_1 l_1}_{i_2 j_2 k_2 l_2}
  ~+~
   c^{i_1 j_1 k_1 k_2}_{i_2 j_2\,l_2\, l_1}
   ~+~
   c^{i_1 j_1 k_1 l_2}_{i_2 j_2\,l_1 k_2}
   ~~=~~ 0 \, .
   \begin{picture}(0,0)(0,0)
   \put(-151.4,11){\line(1,0){7.5}}
   \put(-158.9,2.8){\line(1,0){7.7}}
   \put(-158.9,-4.8){\line(1,0){14.6}}
   \put(-158.9,3){\line(0,-1){8}}
   \put(-151.4,11.1){\line(0,-1){8.1}}
   \put(-144.1,11){\line(0,-1){16}}
   \put(-98.7,11){\line(1,0){8.5}}
   \put(-106.2,2.8){\line(1,0){7.7}}
   \put(-106.2,-4.8){\line(1,0){15.6}}
   \put(-106.2,3){\line(0,-1){8}}
   \put(-98.7,11.1){\line(0,-1){8.1}}
   \put(-90.4,11){\line(0,-1){16}}
   \put(-44.7,11){\line(1,0){8.5}}
   \put(-52.2,2.8){\line(1,0){7.7}}
   \put(-52.2,-4.8){\line(1,0){15.6}}
   \put(-52.2,3){\line(0,-1){8}}
   \put(-44.7,11.1){\line(0,-1){8.1}}
   \put(-36.4,11){\line(0,-1){16}}
   \end{picture}
\ee
Considering all possible 2-planes, it follows that full $SU(4)_R$ invariance requires that this type of cyclic identity holds for any such choice of an upper index and two lower indices (or two upper indices and one lower).

At the general N$^K$MHV level, an analogous argument shows that invariance under infinitesimal $SU(4)_R$ transformation requires the $c$-coefficients to satisfy $(K+1)$-term cyclic identities. At the N$^3$MHV level we find
\be
  \lab{cyc2}
   c\bigg[
   \hspace{-1mm}
   \begin{array}{rrr}
   \scriptstyle l_1\\[-2mm]
   \scriptstyle l_2\\[-2mm]
   .\, .\, \scriptstyle k_3l_3
   \end{array}
   \hspace{-2mm}
   \bigg]
   ~-~
   c\bigg[
   \hspace{-1mm}
   \begin{array}{rrr}
   \scriptstyle l_2\\[-2mm]
   \scriptstyle l_3\\[-2mm]
   .\, .\, \scriptstyle l_1k_3
   \end{array}
   \hspace{-2mm}
   \bigg]
   ~+~
   c\bigg[
   \hspace{-1mm}
   \begin{array}{rrr}
   \scriptstyle l_3\\[-2mm]
   \scriptstyle k_3\\[-2mm]
   .\, .\, \scriptstyle l_2\,l_1
   \end{array}
   \hspace{-2mm}
   \bigg]
   ~-~
   c\bigg[
   \hspace{-1mm}
   \begin{array}{rrr}
   \scriptstyle k_3\\[-2mm]
   \scriptstyle l_1\\[-2mm]
   .\, .\, \scriptstyle l_3\,l_2
   \end{array}
   \hspace{-2mm}
   \bigg]
   ~~=~~0\, .
\ee
The cyclic identities \reef{cyc1} and \reef{cyc2} and their $(K+1)$-term generalizations continue to hold for the set of $c$-coefficients in the superamplitude \reef{NKMHV1} whose indices are in the range from 1 to $n-4$.

We now present some examples of these linear relations.  The first one involves the amplitudes
\bea
  \nonumber
  &&c^{1113}_{2224} ~=~
  -\, A_8( \lambda^{123} \lambda^{123} \lambda^{4} \lambda^{4}
  ++ --  )\, ,~~~~~~~
  c^{1114}_{2232} ~=~
  -A_8(\lambda^{123} \lambda^{124} \lambda^{3} \lambda^{4}
  ++ --  )\, ,~~~~~~~~~\\[1mm]
  &&
  c^{1112}_{2243} ~=~
  -\, A_8( \lambda^{123} \lambda^{124} \lambda^{4} \lambda^{3}
  ++ --  )\,
  \lab{c2A}
\eea
(see \reef{ctoA} for details).
The cyclic identity \reef{cyc1} relates these three amplitudes,  viz.
\be
  \lab{cycID1}
   c^{1113}_{2224}
   ~+~ c^{1114}_{2232}
   ~+~ c^{1112}_{2243}
   ~=~ 0\,.
\ee

\begin{description}
\item{\bf Application: ``Gluon-stripped'' cyclic identities}\\
Since the cyclic identities are a consequence of $SU(4)_R$ invariance
and the four fixed gluons of the algebraic basis are singlets,  the identities are also  valid if one strips off the four gluon states $++--$. In this case we find a linear relation among three MHV 4-point amplitudes, namely
\bea
  \nonumber
  &&
  \hspace{-1.2cm}
  A_4( \lambda^{123} \lambda^{12{\bf 3}} \lambda^{{\bf 4}} \lambda^{{\bf 4}})
  +
  A_4(  \lambda^{123} \lambda^{12{\bf 4}} \lambda^{{\bf 3}} \lambda^{{\bf 4}})
  +
  A_4(  \lambda^{123} \lambda^{12{\bf 4}} \lambda^{{\bf 4}} \lambda^{{\bf 3}})\\[1mm]
  &\hspace{-.3cm}\propto&\hspace{-.2cm}
    \big[- \< 12 \>^3 \<34\>\big]
    +
    \big[+\<12\>^2 \<13\> \< 24\>\big]
    +
    \big[- \<12\>^2 \<14\> \<23\>\big]
  ~~=~~ 0 \, .~~~~~~
  \lab{excyc}
\eea
The terms $[\dots]$ are the spin factors obtained from Nair's generating function \reef{simple}. The sum of spin factors vanishes by the Schouten identity,  which is a curious way to verify a consequence of an internal symmetry. Note that \reef{excyc} is valid at any loop order.

Let us also consider an example of the N$^3$MHV cyclic identity \reef{cyc2}
Starting from
\begin{equation}
c{\Big[
   \hspace{-3mm}
   \raisebox{1pt}{
     $\begin{array}{r}
     \scriptscriptstyle 1112\\[-2mm]
     \scriptscriptstyle3334\\[-2mm]
     \scriptscriptstyle 5656
     \end{array}$
   }
   \hspace{-3mm}
   \Big]}
   =
   - A_{10}( \l^{123} \l^4 \l^{123} \l^4 s^{13} s^{24} + + - -)\,,
\end{equation}
one generates a non-trivial relations between four 10-point amplitudes.  As above, a four term identity must hold for the 6-point NMHV amplitudes obtained by stripping off the gluons $++--$.
We have verified numerically that
\bea
  \nonumber
  0 &=&A_{6}( \l^{123} \l^{\bf 4} \l^{123} \l^{\bf 4} s^{1{\bf 3}} s^{2{\bf 4}} )
  ~+~ A_{6}( \l^{123} \l^{\bf 4} \l^{123} \l^{\bf 4} s^{1{\bf 4}} s^{2{\bf 3}} ) \\[1mm]
  &&
  ~+~ A_{6}( \l^{123} \l^{\bf 4} \l^{123} \l^{\bf 3} s^{1{\bf 4}} s^{2{\bf 4}} )
  ~+~ A_{6}( \l^{123} \l^{\bf 3} \l^{123} \l^{\bf 4} s^{1{\bf 4}} s^{2{\bf 4}} ) \,
  \lab{6ptex}
\eea
indeed holds at tree level.

`Gluon-stripped' relations such as \reef{excyc} and \reef{6ptex} can also be derived directly starting from a
representation of the superamplitudes  $\ca^{\text{N$^K$MHV}}_n$ without the  explicit $\d^{(8)}$-function factor and an unconstrained
Grassmann polynomial of order $4(K+2)$. At the NMHV level,
we would write
\begin{equation}
\ca_n^\text{NMHV}
= \sum w_{[i_A j_A k_A l_A]} \prod_{B=1}^3 \eta_{i_B 1} \eta_{j_B 2} \eta_{k_B 3} \eta_{l_B 4}\, .
\end{equation}
Each of the 12 indices are summed from 1 to $n$.
Infinitesimal $SU(4)_R$ transformations applied to this form of the generating function would imply precisely the cyclic identity \reef{cyc2} for the $w$-coefficients. Eq.~\reef{6ptex} is an example of this identity.
\end{description}

\subsection{Basis amplitudes and Young tableaux}
\lab{secNKMHVa}

Our goal now is to find the algebraic basis of amplitudes that  determines the N$^K$MHV superamplitude in $\cn=4$ SYM.  The representation
 \reef{NKMHV1} is manifestly SUSY invariant, but
the $c$-coefficients are not linearly independent. Rather they are related by cyclic identities such  as \reef{cyc1} and \reef{cyc2}. Thus we must find a way to characterize the independent $c_{[i_A j_A k_A l_A]}$-coefficients, with indices restricted to $1,2,\dots, n-4$, subject to\\[1.2ex]
(1) antisymmetry of columns, \\
(2) column exchange symmetry, and\\
(3) $(K+1)$-term cyclic identities.

Before tackling the full problem, we remind the reader of a seemingly unrelated problem, namely the counting of independent components of the Riemann tensor in $D$ dimensions. We suggestively write the Riemann tensor $R_{\mu\nu\rho\s} = \mathcal{R}^{\mu\rho}_{\nu\s}$. The familiar symmetries of the indices can then be described as the properties (1) and (2) above.   Furthermore, property (3), namely the 3-term cyclic identity, is exactly the first Bianchi identity $R_{\mu\nu\rho\s}+R_{\mu\s\nu\rho}+R_{\mu\rho\s\nu}=0$.  In our  notation it reads
$\mathcal{R}^{\mu\rho}_{\nu\s}
+\mathcal{R}^{\mu\nu}_{\s\rho}
+\mathcal{R}^{\mu\s}_{\rho\nu}=0$.
The symmetries of the Riemann tensor can be encoded in the 4-box quadratic Young diagram with two rows and two columns. The independent components of the Riemann tensor in $D$-dimensions are in one-to-one correspondence with the semi-standard tableaux constructed from this Young diagram using integers from the set $\{1,\dots,D\}$. (We review  Young tableaux in appendix \ref{app17}.)
The number of semi-standard tableaux is equal to the dimension of the irreducible representation of $SU(D)$ corresponding to this Young diagram. This dimension is easily computed using the hook formula \cite{hook} and the answer is that there are $D^2 (D^2-1)/12$ independent components of $R_{\mu\nu\rho\s}$.

The startling similarity to our problem for the $c$-coefficients leads us to expect that the number of independent basis coefficients $c^{i_1 j_1 k_1 l_1}_{i_2 j_2 k_2 l_2}$ of the N$^2$MHV superamplitudes will be given by the dimension of the irreducible representation of $SU(n-4)$ corresponding to the rectangular  Young diagram with $2$ rows and 4 columns. Using the hook formula \cite{hook} this gives
\be\label{N2MHVcount}
  \text{\#(N$^2$MHV basis amplitudes)} ~=~
  {\rm dim_{SU(n-4)}} \hspace{-1mm}
  \begin{array}{l}
  \framebox[3.5mm][c]{\scriptsize \phantom{1}}
  \framebox[3.5mm][c]{\scriptsize \phantom{1}}
  \framebox[3.5mm][c]{\scriptsize \phantom{1}}
  \framebox[3.5mm][c]{\scriptsize \phantom{1}}\\[-2pt]
  \framebox[3.5mm][c]{\scriptsize \phantom{1}}
  \framebox[3.5mm][c]{\scriptsize \phantom{1}}
  \framebox[3.5mm][c]{\scriptsize \phantom{1}}
  \framebox[3.5mm][c]{\scriptsize \phantom{1}}
  \end{array}
  ~=~
  \frac{(n-5)(n-4)^2(n-3)^2(n-2)^2(n-1)}{4! \, 5!} \, .
\ee

It is clear to the eye that the index pattern of $c^{i_1 j_1 k_1 l_1}_{i_2 j_2 k_2 l_2}$ matches precisely to fillings of this rectangular Young diagram. The antisymmetry of the columns also matches the requirements of the Young diagram.  We now argue that the independent  coefficients (and thus
the independent basis amplitudes) are precisely those whose indices correspond
to  semi-standard tableaux of this  Young diagram.

For general $K$ we will show that the amplitudes in the algebraic basis are those whose $4 K$ indices form a
 $SU(n-4)$ semi-standard tableau of a rectangular
Young diagram with $K$ rows and $4$ columns. To show this, we proceed in three steps. First, we count the total number of $SU(4)_R$ singlets that can be formed from the amplitudes encoded in the $c$-coefficients. Then we connect this counting of singlets to the dimension of the $SU(n-4)$ representation associated with the $K$-by-$4$ Young diagram. Finally, we show how arbitrary
$c$-coefficients can be expressed in terms of basis coefficients.

\vspace{2mm}
\noindent {\bf Counting $SU(4)_R$ singlets}\\
Recall that
$K$ sets of $SU(4)_R$ indices must be distributed among the $n-4$ particle states.
 These distributions are represented by the indices of $c_{[i_A, j_A, k_A, l_A]}$.  As described in Section \ref{secAc}, these indices  simply tell us that states $i_1,\dots, i_K$ carry $SU(4)_R$ index 1,  states $j_1,\dots, j_K$ carry $SU(4)_R$ index 2, etc. Each state can carry $w=0, 1, 2, 3$,  or 4 $SU(4)_R$ indices, with $0$ corresponding to a positive-helicity gluon, $1$ is a positive-helicity gluino etc. Thus any partition $\lambda= [w_1,w_2,\dots,w_{n-4}]$ of $4K$, with $0\!\leq w_i\!\leq\! 4$, defines a specific set of external particle types. Distinct assignments of  these particle types to the external lines give
algebraically independent amplitudes.  For a given partition $\lambda$, the number of distinct
 assignments is given by the multinomial coefficient $\mathcal{C}_\l$ defined in \reef{Cl}.

$SU(4)_R$ transformations do not change the particle types nor their order within an amplitude. Instead they reshuffle the $SU(4)_R$ indices. Thus the cyclic identities relate amplitudes with the same
 assignment of particle types, i.e.~within a given partition $\lambda$ of $4K$, with a fixed ordering. This is illustrated in the examples \reef{excyc} and \reef{6ptex}. The particles  of the $\cn=4$ theory transform as fully antisymmetric representations of $SU(4)_R$: the number of boxes in the corresponding
 single-column Young diagram is simply the number of indices carried by the corresponding particle. \emph{For a given partition $\lambda$,
 the number of  independent amplitudes obtained from assigning
 $SU(4)_R$ indices to the set of particles
is equal to the number of singlets $\cs_\lambda$ in the decomposition of the product of  $SU(4)_R$ irreps corresponding to the $n-4$ external states.}
The point is that these singlets are
 R-invariant by definition,
so they cannot be related to each other by the cyclic identities, which arose from requiring R-symmetry.

Thus, for a given partition $\lambda$, the total number of independent amplitudes is $\mathcal{C}_\l \mathcal{S}_\l$. We must consider all partitions $\lambda= [w_1,w_2,\dots,w_{n-4}]$ of $4K$ with $0 \le w_i \le 4$. The total number of independent amplitudes will therefore be
$\sum_{\l} \mathcal{C}_\l \mathcal{S}_\l$.

Let us demonstrate this in a few examples. For NMHV amplitudes,
there is only a single set of $SU(4)_R$ indices to distribute among $n-4$ lines. Thus the singlet count for any partition is $\mathcal{S}_\lambda =1$, and the count of basis amplitudes is simply $\sum_{\lambda}\mathcal{C}_\l$. This is also the number of independent components of a fully symmetric 4-index tensor.
For $n=6$, the partition $\l=[4,0]$ has $\mathcal{C}_{[4,0]}=2$, and we find $\mathcal{C}_{[3,1]}=2$ and $\mathcal{C}_{[2,2]}=1$. The number of  6-point basis amplitudes is therefore $\sum_{\lambda}\mathcal{C}_\l=2+2+1=5$\,, which reproduces our result from Section \ref{secEx2}.

\begin{table}
\begin{displaymath}
\begin{array}{crcr}
 \lambda & \mathcal{C}_\l ~~~~&\mathcal{S}_\l~~
 &\mathcal{C}_\l \mathcal{S}_\l ~\\
 {}[4, 4, 0, 0] & 6 ~~~~~& 1 ~~~& 6~~~ \\
 {}[4, 3, 1, 0] & 24 ~~~~~& 1 ~~~& 24~~~\\
 {}[4, 2, 2, 0] & 12 ~~~~~& 1 ~~~& 12~~~\\
 {}[4, 2, 1, 1] & 12 ~~~~~& 1 ~~~& 12~~~\\
 {}[3, 3, 2, 0] & 12 ~~~~~& 1 ~~~& 12~~~\\
 {}[3, 3, 1, 1] & 6 ~~~~~& 2 ~~~& 12~~~\\
 {}[3, 2, 2, 1] & 12 ~~~~~& 2 ~~~& 24~~~\\
 {}[2, 2, 2, 2] & 1 ~~~~~& 3 ~~~& 3~~~\\
 \hline
  8~\text{partitions}
  & \sum_{\lambda} \mathcal{C}_\l \mathcal{S}_\l
  &=~& 105~~~
\end{array}
\hspace{1cm}
\begin{array}{crcr}
 \lambda~~ & \mathcal{C}_\l ~~~~&\mathcal{S}_\l~~
 &\mathcal{C}_\l \mathcal{S}_\l ~\\
 {}[4, 4, 0, 0, 0] ~~& 10~~~~ & 1 ~~~& 10 ~~~ \\
 {}[4, 3, 1, 0, 0] ~~& 60~~~~ & 1 ~~~& 60 ~~~ \\
 {}[4, 2, 2, 0, 0] ~~& 30~~~~ & 1 ~~~& 30 ~~~ \\
 {}[4, 2, 1, 1, 0] ~~& 60~~~~ & 1 ~~~& 60 ~~~ \\
 {}[4, 1, 1, 1, 1] ~~& 5~~~~ & 1 ~~~& 5 ~~~ \\
 {}[3, 3, 2, 0, 0] ~~& 30~~~~ & 1 ~~~& 30 ~~~ \\
 {}[3, 3, 1, 1, 0] ~~& 30~~~~ & 2 ~~~& 60 ~~~\\
 {}[3, 2, 2, 1, 0] ~~& 60~~~~ & 2 ~~~& 120 ~~~\\
 {}[3, 2, 1, 1, 1] ~~& 20~~~~ & 3 ~~~& 60 ~~~\\
 {}[2, 2, 2, 2, 0] ~~& 5~~~~ & 3 ~~~& 15 ~~~\\
 {}[2, 2, 2, 1, 1] ~~& 10~~~~ & 4 ~~~& 40 ~~~\\
  \hline
  11~\text{partitions}
  & \sum_{\lambda} \mathcal{C}_\l \mathcal{S}_\l
  &=~& 490~~~
\end{array}
\end{displaymath}
\caption{Partitions $\l$ of $\cn K = 8$ of length $n-4$ for $n=8$ (left) and $n=9$ (right).
The multinomial coefficient $\mathcal{C}_\l$ and the
 number of singlets $\mathcal{S}_\l$ are displayed for each partition. The sum  $\sum_{\lambda} \mathcal{C}_\l \mathcal{S}_\l$ equals the dimension of the $SU(n-4)$ irrep corresponding to the 2-by-4 rectangular Young diagram.}
\label{tab0}
\end{table}

The first non-trivial N$^2$MHV amplitudes have 8 or 9 external states. In Table \ref{tab0} we tabulate the combinatorial factor $\mathcal{C}_\l$ and singlet count $\mathcal{S}_\l$ for each partition.
As an example, consider the $n=9$ partition $\l=[2,2,2,2,0]$.
In $SU(4)_R$ the decomposition of the product
\be
 \begin{array}{l}
  \framebox[3.5mm][c]{\scriptsize \phantom{1}}\\[-2pt]
  \framebox[3.5mm][c]{\scriptsize \phantom{1}}
  \end{array}
  \otimes
  \begin{array}{l}
  \framebox[3.5mm][c]{\scriptsize \phantom{1}}\\[-2pt]
  \framebox[3.5mm][c]{\scriptsize \phantom{1}}
  \end{array}
  \otimes
  \begin{array}{l}
  \framebox[3.5mm][c]{\scriptsize \phantom{1}}\\[-2pt]
  \framebox[3.5mm][c]{\scriptsize \phantom{1}}
  \end{array}
  \otimes
  \begin{array}{l}
  \framebox[3.5mm][c]{\scriptsize \phantom{1}}\\[-2pt]
  \framebox[3.5mm][c]{\scriptsize \phantom{1}}
  \end{array}
  \otimes 1
\ee
contains 3 singlets, hence $\mathcal{S}_\l=3$. In other words, there are 3 inequivalent ways  $SU(4)_R$ indices can be placed on the four scalars $s^{ab}$ in the basis amplitudes $A_9(s\,s\,s\,s\,+++--)$. These are
\bea
  \nonumber
  &&c^{1122}_{3344} = A_9(s^{12} s^{34} s^{12} s^{34}+++--)\, ,~~~~~~~
  c^{1123}_{2344} = A_9(s^{12} s^{13} s^{24} s^{34}+++--)\, ,~~\\[1mm]
  &&c^{1133}_{2244} = A_9(s^{12} s^{12} s^{34} s^{34}+++--)\, .
\eea
The multinomial coefficient $\mathcal{C}_\l = 5$ counts the five different placements of the positive-helicity gluon among the five first lines. The total number of basis amplitudes associated with the partition $\l=[2,2,2,2,0]$ is then $\mathcal{C}_\l \mathcal{S}_\l =15$.
This singlet count precisely reproduces the counting of semi-standard Young tableaux in~(\ref{N2MHVcount}).
 We now show that these approaches are indeed equivalent.

\vspace{2mm}
\noindent {\bf Relating $SU(4)_R$ singlets to $SU(n-4)$ Young tableaux}\\
Consider the semi-standard $SU(n-4)$ tableaux of the rectangular $K$-by-4 Young diagram,
which are obtained by filling in numbers from the set $\{1,2,\dots,n-4\}$. The
 required concepts are introduced in appendix~\ref{app17}. The multiplicities of each entry in
 a semi-standard tableau form a partition of $4K$. For any partition $\lambda= [w_1,w_2,\dots,w_{n-4}]$ of $4K$, the number of distinct ways one can order the $w_i$'s to assign weights
 to a semi-standard tableau is simply given by the multinomial coefficient $\mathcal{C}_\l$. For each such weight assignment there are $\tilde{\mathcal{S}}_\l$ distinct semi-standard tableaux. The number $\tilde{\mathcal{S}}_\l$ is
 called the \emph{Kostka number}.
 It is independent of the ordering of weights and depends only on the partition $\lambda$ \cite{fulton}. We show in appendix
\ref{app17} that $\tilde{\mathcal{S}}_\l$ is equal to the singlet count
$\mathcal{S}_\l$ introduced above; $\tilde{\mathcal{S}}_\l = \mathcal{S}_\l $. Thus the number of semi-standard tableaux with the weights of a given partition $\l$ is $\mathcal{C}_\l \mathcal{S}_\l$. The total number of semi-standard tableaux $\sum_{\l} \mathcal{C}_\l \mathcal{S}_\l$ is equal to the dimension $d_{Y}$ of the $SU(n-4)$ irrep corresponding to the rectangular $K$-by-4 Young diagram $Y$. This proves the claim above that the number of amplitudes in the algebraic basis of N$^K$MHV equals $d_Y$.

\vspace{2mm}
\noindent {\bf Expressing non-basis amplitudes in the algebraic basis}\\
The result that the singlet count $\mathcal{S}_\l$ is equal to the Kostka number, $\mathcal{S}_\l= \tilde{\mathcal{S}}_\l$,
 suggests that the basis amplitudes are labeled by semi-standard tableaux.
For a given partition $\lambda$, any set of indices of $c_{[i_A j_A k_A l_A]}$ that does
\emph{not} form a semi-standard tableau can
 indeed be expressed as a linear combination of those that do. The procedure --- called  the \emph{straightening process} \cite{fulton} --- involves repeated use of
the cyclic identities.
For $K=2$ we implemented this in Mathematica, but the
Macaulay2 code
 of \cite{maccy} can also be used.
A simple example of straightening is
\be
   c^{1113}_{6425}
   ~=~
   c^{1113}_{2465}
   ~=~
   c^{1113}_{2456}
   -
   c^{1115}_{2346}\,.
\ee
The first step uses the column exchange symmetry, and the second step is an application of the cyclic identity \reef{cyc1} and of antisymmetry within columns.
Generally, several applications of the cyclic identity may be needed. For example, after several steps one obtains
\be
  c^{1235}_{6487}
  ~=~
  c^{1235}_{4678}
  -
  c^{1345}_{2678}
  +
  c^{1356}_{2478}
  -
  2 \,  c^{1357}_{2468}
  -
  c^{1237}_{4568}
  +
  c^{1347}_{2568}
  +
  c^{1257}_{3468} \, .
\ee

To express a general $c$-coefficient in terms of basis coefficients, one proceeds as follows. One first uses column-exchange symmetry to completely order the first row, and to order the remaining rows  as much as possible. One then uses the cyclic identity on the first pair of columns that violates the semi-standard filling. These two steps are iterated until only basis coefficients remain. One may worry that the repeated use of cyclic identities could continue in endless loops. However, the process does stop in a finite number of steps. For the N$^2$MHV case one can see this from the fact that the function $f = i_1 +j_1 +k_1 + l_1$ increases
at each application of the cyclic identity and is bounded
 from above.

The inverse straightening process gives the set of non-semi-standard tableaux whose
 ``straightened'' expressions contain a given semi-standard tableau. We have implemented this process in a Mathematica code for the N$^2$MHV case and
 it guided us in the construction of a manifestly $SU(4)_R$-invariant form of the N$^2$MHV
superamplitude, which we present in the next section.

Before we end this section, let us summarize the results.
We have shown that
\begin{quote}
\sl
The algebraic basis of amplitudes for the N$^K$MHV sector of $\cn=4$ SYM is given by the amplitudes associated with $c$-coefficients whose indices $[i_A,j_A,k_A,l_A]$ run over
 external lines $1,2,\dots,n-4$ and form a semi-standard tableau of the rectangular $K$-by-4 Young diagram. The number of basis amplitudes is therefore the dimension of the $SU(n-4)$ irrep corresponding to this Young diagram.
\end{quote}

For $K=1$ this agrees with our NMHV results of Section \ref{secNMHV}. In that case the basis amplitudes were labeled by the fully symmetric $c_{ijkl}$-coefficient, whose independent components exactly map to the semi-standard tableaux of the 4-box single row Young diagram.

Any N$^K$MHV $n$-point amplitude also has an interpretation as an anti-N$^{(n-4-K)}$MHV $n$-point amplitude. Therefore our dimension formula must give the same result
 when $K \to n-4-K$. To see that this works, recall that in $SU(n-4)$ the conjugate representation of the irrep corresponding to the $K$-by-4 Young diagram is an irrep whose Young diagram has $n-4-K$ rows and 4 columns.
Since conjugate representations have the same dimension, our prescription automatically incorporates the fact that N$^K$MHV $n$-point amplitudes can also be described as anti-N$^{(n-4-K)}$MHV amplitudes.

\subsection{N$^2$MHV superamplitude}
\lab{secN2MHVb}

The superamplitude  \reef{NKMHV1} is expressed in terms of the SUSY-invariant polynomials
 $X_{[i_A,j_A,k_A,l_A]}$  given in \reef{genX1}, which we simply denote by $X^{i_1 j_1 k_1 l_1}_{\hspace{1pt}i_ 2j_ 2k_2 l_2}$ in the N$^2$MHV sector. In the previous section, we showed that these polynomials transform non-trivially under infinitesimal R-symmetry transformations. The requirement that the full superamplitude is $SU(4)_R$ invariant therefore imposes a set of cyclic
 identities~(\ref{cyc1}) that relate the amplitude coefficients $c^{i_1 j_1 k_1 l_1}_{i_ 2j_ 2k_2 l_2}$. As discussed above,
the independent basis amplitudes are precisely the amplitudes
 $A_n\bigl(\,
\bigl\{
  {}^{i_1 j_1  k_1  l_1}_{i_2 j_2  k_2  l_2}
\bigr\}
\,++--\,\bigr)$
 corresponding to semi-standard tableaux of $SU(n-4)$.  We now wish to write the superamplitude in terms of these basis amplitudes only, viz.
\begin{equation}
  \lab{N2MHV1}
   \ca^{\text{N$^2$MHV}}_n ~=~
   \frac{1}{16}
     \sum_{{}^\text{semi-standard}_\text{~~tableaux Y}}\!\!\! (-)^Y
     A_n\bigl(\,
     \bigl\{
    {}^{i_1 j_1  k_1  l_1}_{i_2 j_2  k_2  l_2}
    \bigr\}
     ++--\bigr)~
      Z^{i_1 j_1 k_1 l_1}_{\hspace{1pt}i_ 2j_ 2k_2 l_2}
     \,,
\end{equation}
The sum over semi-standard tableaux is equivalent to the requirement that the indices satisfy $i_A \le j_A \le k_A \le l_A$ for $A=1,2$ and that each column is ordered so that the smaller number appears in the first row, $i_1 < i_2$ etc. The $(-)^Y$ is the sign factor $\pm1$ that was explained in Section \ref{secAc}.

It remains to determine the $\eta$-polynomials $Z^{i_1 j_1 k_1 l_1}_{\hspace{1pt}i_ 2j_ 2k_2 l_2}$ of degree $12$ in~(\ref{N2MHV1}). The polynomials
$Z^{i_1 j_1 k_1 l_1}_{\hspace{1pt}i_ 2j_ 2k_2 l_2}$ must be $SU(4)_R$-invariant; if not, invariance would impose further linear relations among the basis
amplitudes. Starting with $X^{i_1 j_1 k_1 l_1}_{\hspace{1pt}i_ 2j_ 2k_2 l_2}$, is it easy to see that if we sum all
 distinct permutations of indices in each row, then the result is an $SU(4)_R$-invariant, $X^{(i_1 j_1 k_1 l_1)}_{\hspace{1pt}(i_ 2j_ 2k_2 l_2)}$. However, in some cases the symmetrization includes $X$-polynomials of other semi-standard tableaux; these would prevent  the correct extraction of the pure basis amplitudes from the superamplitude, so we must remove them in an $SU(4)_R$-invariant way.

As an example, consider $X^{1247}_{\hspace{1pt}3568}$. Its symmetrized form, $X^{(1247)}_{\hspace{1pt}(3568)}$, includes the term
$X^{1274}_{\hspace{1pt}5638}=-X^{1234}_{\hspace{1pt}5678}$ whose indices form a different semi-standard tableau. We remove its symmetrized form,
 $X^{(1234)}_{\hspace{1pt}(5678)}$,
which does not contain any further new semi-standard tableaux. Thus we have found that the SUSY and R-symmetry invariant $\eta$-polynomial  multiplying the basis
 amplitude $A_n\bigl(\,\bigl\{{}^{1247}_{3568} \bigr\}++--\bigr)$ is
\begin{equation}
    Z^{1247}_{\hspace{1pt}3568} = X^{\hspace{1pt}(1247)}_{\hspace{1pt}(3568)} + X^{(1234)}_{\hspace{1pt}(5678)}\,.
\end{equation}

We now set up a systematic recursive procedure to determine the desired
$Z^{i_1 j_1 k_1 l_1}_{\hspace{1pt}i_ 2j_ 2k_2 l_2}$.
We define
\be\label{defZs}
 Z^{i_1 j_1 k_1 l_1}_{\hspace{1pt}i_ 2j_ 2k_2 l_2}~=~X^{i_1 j_1 k_1 l_1}_{\hspace{1pt}i_ 2j_ 2k_2 l_2} ~
  + \bigl[\,W^{(i_1 j_1k_1 l_1)}_{ \,(i_ 2 j_ 2 k_2 l_2)}-W^{i_1 j_1 k_1 l_1}_{\,i_ 2j_ 2k_2 l_2}\,\bigr]\, ,
\ee
where for canonical ordering
$i_1<i_2,~j_1<j_2,$ etc.
\be\lab{defWs}
W^{i_1 j_1 k_1 l_1}_{i_ 2j_ 2k_2 l_2} \equiv
\bigg\{
\begin{array}{lcl}
X^{i_1 j_1 k_1 l_1}_{\hspace{1pt}i_ 2j_ 2k_2 l_2} - Z^{i_1 j_1 k_1 l_1}_{\hspace{1pt}i_ 2j_ 2k_2 l_2}
&&\text{if indices form semi-standard tableau}\,,\\[1mm]
X^{i_1 j_1 k_1 l_1}_{\hspace{1pt}i_ 2j_ 2k_2 l_2}  &
&\text{otherwise}.
\end{array}
\ee
The $W$'s are antisymmetric
within each column, just like the $X$'s.
In \reef{defZs}, $W^{(i_1 j_1k_1 l_1)}_{ \,(i_ 2 j_ 2 k_2 l_2)}$ is the sum over all distinct permutations of each row $i_1 j_1k_1 l_1$ and $i_ 2j_ 2k_2 l_2$, and the term $-W^{i_1 j_1 k_1 l_1}_{\,i_ 2j_ 2k_2 l_2}$ excludes the identity permutation from this sum.

The reader may worry that the recursive approach \reef{defZs}-\reef{defWs} runs in endless circles. However, one should note that new semi-standard tableaux only appear after symmetrizations if one or more
 ``column flips'' are performed on the indices. Since each column flip brings a smaller integer from the lower row to the top row, it reduces the sum of the indices in the top row. Since this sum is bounded from the below, the process will eventually stop. With $Z$-polynomials defined this way, the N$^2$MHV superamplitude \reef{N2MHV1} is manifestly SUSY and R-symmetry invariant, and correctly expresses arbitrary N$^2$MHV amplitudes in terms of the algebraic basis amplitudes.

\subsection{Basis amplitudes in $\cn = 8$ supergravity}
\lab{sec8}
In supergravity, the analysis of SUSY Ward identities and $SU(8)_R$ symmetries is carried out the same way as in SYM theory and leads to
the \emph{algebraic} basis for amplitudes and superamplitudes at the N$^K$MHV level.   The basis amplitudes correspond to semi-standard tableaux
of Young diagrams with 8 columns and $K$ rows.
As in Section \ref{secEx2},  many amplitudes of the algebraic basis are related by momentum relabeling, and it is important to study these relations
because they yield a much smaller basis of  \emph{functionally independent} amplitudes.   This task requires two stages. The first stage
can be carried out as a systematic  group theory analysis and leads to a large reduction of the basis.  The second stage is needed to identify further functional relations due to a combination of momentum relabeling and cyclic identities.  We identify the mechanism  for these relations in an example, but stop short of a complete analysis.  As a preview we state results for the N$^K$MHV 8-point superamplitude.  There are 825  amplitudes in the algebraic basis; the first set of functional relations gives a reduction to 63, and  using cyclic relations we were able to further reduce this to a set  of 46 functional basis amplitudes.

 \paragraph{The algebraic basis.} The N$^K$MHV superamplitudes in $\cn =8$ supergravity are  degree $8(K+2)$ Grassmann polynomials. The algebraic basis consists of amplitudes with states $n-2$ and $n-3$ fixed to be positive-helicity gravitons and states $n-1$ and $n$ fixed to be negative-helicity gravitons. The remaining $n-4$ states are determined by the indices of the corresponding $c$-coefficients.  The basis amplitudes carry the index structure of semi-standard tableaux of the Young diagram with $K$ rows and $8$ columns.  As in \reef{ctoA}, the $c$-coefficients are equal to specific amplitudes up to a sign factor.

 \noindent For example, the N$^2$MHV superamplitude takes the form
\begin{equation}
  \lab{sgN2MHV1}
   \cm^{\text{N$^2$MHV}}_n ~=~
   \frac{1}{256}
     \sum_{{}^\text{semi-standard}_\text{~~tableaux Y}}\!\!\! (-)^Y
     M_n
     \bigl(\,
     \bigl\{
    {}^{i_1 j_1k_1 l_1 p_1 q_1 u_1 v_1}_{i_2 j_2 k_2 l_2 p_2 q_2 u_2v_2}     \bigr\}
     ++--\bigr)~
      Z^{i_1 j_1k_1 l_1 p_1 q_1 u_1 v_1}_{\hspace{1pt}i_2 j_2  k_2 l_2 p_2 q_2 u_2 v_2}
     \,,
\end{equation}
where the SUSY and $SU(8)_R$-invariant $\eta$-polynomials $Z$ are defined in complete analogy to
 the gauge theory expression \reef{defZs}.

\paragraph{Functional relations --first stage.}
Several amplitudes in the algebraic basis are functionally related by permutations of the external momentum labels. We discussed this in Section~\ref{secEx2} for the NMHV sector. To count the number of functionally independent basis amplitudes at level  N$^K$MHV,  we review the process that let us to characterize the basis in terms of semi-standard tableaux. First, consider a
 fixed partition $\l=[w_1,w_2,\ldots,w_{n\dash4}]$ of $8K$ with $0\!\leq\! w_i\! \leq\! 8$.  It determines a specific set of external particles. The Kostka number $\mathcal{S}_\l$ is the number of independent ways $SU(8)_R$ indices can be distributed on the particles states specified  by $\l$.  The ordering of these states has no meaning in gravity, so  in the count of functionally independent amplitudes  we do \emph{not} include the multinomial coefficient $\mathcal{C}_\l$ (which \emph{is} required in $\cn =4$ SYM). We  conclude from this that  there are
 at most $s = \sum_\l \mathcal{S}_\l$ functionally independent basis amplitudes. However, as stated above,  there are further reductions, so this first stage result
is only an upper bound on the functional basis.

\noindent Let us list a few examples of this upper bound at the N$^2$MHV level
\be  \lab{bcn2}
 \begin{array}{rccccccc}
 n~=  & 6 & 7 & 8& 9 & 10 & 11 \\[1ex]
 \text{basis count}\leq s~= & 1 &10 & 63 & 210 & 524 & 1021
 \end{array}\,,
\ee
and at the N$^3$MHV level:
\be \lab{bcn3}
 \begin{array}{rccccccc}
 n~= & 7 & 8 & 9 & 10 & 11 \\[1ex]
 \text{basis count}\leq s~= & 1 & 15 & 210  & 1732 & 8752
 \end{array}
\ee

\begin{table}
\begin{displaymath}
\begin{array}{rrcr}
 \lambda~~~~~~~ & \mathcal{C}_\l ~~~~&\mathcal{S}_\l~~
 &\mathcal{C}_\l \mathcal{S}_\l ~\\
 {}[8, 8, 0, 0] ~~& 6 ~~~~~& 1~~& 6~~ \\
 {}[8, 7, 1, 0] ~~& 24 ~~~~~& 1~~& 24~~\\
 {}[8, 6, 2, 0] ~~& 24 ~~~~~& 1~~& 24~~\\
 {}[8, 6, 1, 1] ~~& 12 ~~~~~& 1~~&12~~\\
 {}[8, 5, 3, 0] ~~& 24 ~~~~~& 1~~& 24~~\\
 {}[8, 5, 2, 1] ~~& 24 ~~~~~& 1~~& 24~~\\
 {}[8, 4, 4, 0] ~~& 12 ~~~~~& 1~~& 12~~\\
 {}[8, 4, 3, 1] ~~& 24 ~~~~~& 1~~& 24~~\\
 {}[8, 4, 2, 2] ~~& 12 ~~~~~& 1~~& 12~~\\
 {}[8, 3, 3, 2] ~~& 12 ~~~~~& 1~~& 12~~\\
 {}[7, 7, 2, 0] ~~& 12 ~~~~~& 1~~& 12~~\\
 {}\bullet[7, 7, 1, 1] ~~& 6 ~~~~~& 2~~& 12~~\\
 {}[7, 6, 3, 0] ~~& 24 ~~~~~& 1~~& 24~~\\
 {}[7, 6, 2, 1] ~~& 24 ~~~~~& 2~~& 48~~\\
 {}[7, 5, 4, 0] ~~& 24 ~~~~~& 1~~& 24~~\\
 {}[7, 5, 3, 1] ~~& 24 ~~~~~& 2~~& 48~~\\
 {}\bullet[7, 5, 2, 2] ~~&~~~ 12 ~~~~~& 2~~& 24~~
\end{array}
\hspace{1.5cm}
\begin{array}{rrcr}
 \lambda~~~~~~~~~ & \mathcal{C}_\l ~~~~&\mathcal{S}_\l~~
 &\mathcal{C}_\l \mathcal{S}_\l ~\\
 {}\bullet[7, 4, 4, 1] ~~~~& 12 ~~~~~& 2~~& 24~~\\
 {}[7, 4, 3, 2] ~~~~& 24 ~~~~~& 2~~& 48~~\\
 {}\bullet[7, 3, 3, 3] ~~~~& 4 ~~~~~& 2~~& 8~~\\
 {}[6, 6, 4, 0] ~~~~& 12 ~~~~~& 1~~& 12~~\\
 {}\bullet[6, 6, 3, 1] ~~~~& 12 ~~~~~& 2~~& 24~~\\
 {}\bullet[6, 6, 2, 2] ~~~~& 6 ~~~~~& 3~~& 18~~\\
 {}[6, 5, 5, 0] ~~~~& 12 ~~~~~& 1~~& 12~~\\
 {}[6, 5, 4, 1] ~~~~& 24 ~~~~~& 2~~& 48~~\\
 {}[6, 5, 3, 2] ~~~~& 24 ~~~~~& 3~~& 72~~\\
 {}\bullet[6, 4, 4, 2] ~~~~& 12 ~~~~~& 3~~& 36~~\\
 {}\bullet[6, 4, 3, 3] ~~~~& 12 ~~~~~& 3~~& 36~~\\
 {}\bullet[5, 5, 5, 1] ~~~~& 4 ~~~~~& 2~~& 8~~\\
 {}\bullet[5, 5, 4, 2] ~~~~& 12 ~~~~~& 3~~& 36~~\\
 {}\bullet[5, 5, 3, 3] ~~~~& 6 ~~~~~& 4~~& 24~~\\
 {}\bullet[5, 4, 4, 3] ~~~~& 12 ~~~~~& 4~~& 48~~\\
 {}\bullet[4, 4, 4, 4] ~~~~& 1 ~~~~~& 5~~& 5~~\\
  \hline
  33 ~\text{partitions}
  &  & \!\!\!\!\!\!\!\!\!\!\!\!\!\!\sum_{\lambda}  \mathcal{S}_\l=63~~~~~~~~\; & \!\!\!\!\!\!\!\!\!\!\!\!\!\!\!\sum_{\lambda}\!  \mathcal{C}_\l\mathcal{S}_\l\!\!=\! 825~~
\end{array}
\end{displaymath}
\caption{Counting independent basis amplitudes for N$^2$MHV 8-point amplitudes of $\cn=8$ supergravity. The table lists the
 Kostka number $\mathcal{S}_\l$ for each of the 33 partitions of 16 of length 4.
 The $\bullet$ marks partitions for which $\mathcal{S}_\l$ may overcount the number of functionally independent amplitudes.}
\label{tab1}
\end{table}

\noindent  A detailed example of this counting is given in Table 1 for N$^2$MHV 8-point amplitudes. There are 33 partitions of
 $8K=16$
of length $n-4=4$. The
  Kostka number $\mathcal{S}_\l$ is listed for each partition,
and  $s = \sum_\l \mathcal{S}_\l=63$. We also list the multinomial factors $\mathcal{C}_\l$ so that the significant reduction due to non-ordering of the states is clear; ordering would have yielded a basis of 825 amplitudes rather than just 63.
The entries in Table 1 were obtained using Mathematica code to compute  $\mathcal{C_\l}$ and
$S_\l$ for each partition, and the same program was used to compute the basis count
in \reef{bcn2} and \reef{bcn3}.

\paragraph{Functional relations - second stage.} However, the first stage result  $s = \sum_\l \mathcal{S}_\l$ actually overcounts the number of functionally independent basis functions. To see this in the 8-point N$^K$MHV example, consider the partition $\lambda=[7,7,1,1]$. The  Kostka number is $\mathcal{S}_{[7,7,1,1]} = 2$ and the corresponding amplitudes have four external gravitinos
\bea
  \nonumber
  c^{11111112}_{22222234} &=&
 M_8\bigl(\,
     \bigl\{
    {}^{11111112}_{22222234}
    \bigr\}
     ++--\bigr)
  ~=~
  M_8\bigl(\,
    \psi^{1234567} \psi^{123456{\bf 8}} \psi^{\bf 7} \psi^{\bf 8}
     ++--\bigr) \, ,\\[1mm]
  c^{11111113}_{22222224} &=&
  -M_8\bigl(\,
     \bigl\{
    {}^{11111113}_{22222224}
    \bigr\}
     ++--\bigr)
     ~=~
    -M_8\bigl(\,
    \psi^{1234567} \psi^{123456{\bf 7}} \psi^{\bf 8}  \psi^{\bf 8}
     ++--\bigr) \, .~~~~
     \lab{A8s}
\eea
If we apply the cyclic identity to the second amplitude, the 3 boldface indices will be shuffled cyclically,
$c^{11111113}_{22222224}~=~
c^{11111112}_{22222234}~-~
c^{11111112}_{22222243}$. The second set of indices form a non-semi-standard tableau, but the corresponding amplitude is simply related to the first amplitude \reef{A8s} by relabeling the momenta 3 and 4, viz.\
\be
  M_8\bigl(\,\psi^{1234567} \psi^{123456{\bf 7}} \psi^{\bf 8}  \psi^{\bf 8} ++--\bigr)
  ~=~
  -M_8\bigl(\,\psi^{1234567} \psi^{123456{\bf 8}} \psi^{\bf 7} \psi^{\bf 8} ++--\bigr) ~+~ (3 \lra 4)\,.~~~~~~~
  \lab{cycA8}
\ee
Thus the two amplitudes of \reef{A8s} are actually functionally dependent. Note that this functional dependence allows us to express the second amplitude in~(\ref{A8s})
in terms of the first, but not vice versa.

For some partitions $\lambda$, relations between the $\mathcal{S}_\l$ amplitudes can
 be found even   without resorting to cyclic identities. Consider for example the two $8$-point amplitudes
\bea
  \nonumber
  c^{11122222}_{22333444} &=&
 \,M_8\bigl(\,
     \bigl\{
    {}^{11122222}_{22333444}
    \bigr\}
     ++--\bigr)
  ~=~
  \,\,M_8\bigl(\,
    \chi^{123} \psi^{1245678} \chi^{345} \chi^{678}
     ++--\bigr) \, ,\\[1mm]
  c^{11122223}_{22233444} &=&
  \!\!\!-M_8\bigl(\,
     \bigl\{
    {}^{11122223}_{22233444}
    \bigr\}
     ++--\bigr)
     \!~=~
    \!\!\!-M_8\bigl(\,
        \chi^{123} \psi^{1234567}\chi^{458} \chi^{678}
     ++--\bigr) \, .~~~~
     \lab{A8s2}
\eea
The coefficients are  related by the momentum relabeling $1\leftrightarrow 4$:
\begin{equation}
    c^{11122222}_{22333444} ~~\to~~ c^{44422222}_{22333111} ~=~ c^{11122223}_{22233444}\,.
\end{equation}
This translates into the amplitude relation
\begin{equation}\label{simplerel}
    M_8\bigl(\,
    \chi^{123} \psi^{1245678} \chi^{345} \chi^{678}
     ++--\bigr)
    ~=~ -M_8\bigl(\,
        \chi^{123} \psi^{1234567} \chi^{458} \chi^{678}
     ++--\bigr)\,.
\end{equation}

\noindent The examples show that $s = \sum_\l \mathcal{S}_\l$ overcounts the number of functionally independent basis amplitudes. Note that an overcount can only occur if at least two lines  $i$, $j$ carry the \emph{same non-graviton} particle, \ie if two weights $w_i$, $w_j$ in the partition $\lambda$ satisfy $1\!\leq\! w_i\!=\!w_j\!\leq\!7$. For example, the singlet count for $\lambda=[7,6,2,1]$ is truly 2; since all particle types are in distinct  $SU(8)_R$ irreps, momentum relabeling cannot give any further relations between the two basis amplitudes.

\noindent In Table 1 we marked a $\bullet$
 next to all partitions for which $\mathcal{S}_\l$ may overcount the functionally independent amplitudes. The largest reduction would leave one independent amplitude in each case. Thus the maximal possible reduction for this example is to count just 1 instead of $\mathcal{S}_\l$ for each $\bullet$'ed partition; the result is 39. The true count $d$ of functionally independent basis amplitudes is therefore $39 \le d \le 63$.  Examining each partition, we found relations of the types~(\ref{cycA8}) and~(\ref{simplerel}) between the $63$ amplitudes and were able to reduce the number of functionally independent amplitudes to $d=46$.
A systematic analysis of the residual functional dependence within each partition $\lambda$ would be interesting and could reveal a general combinatoric structure.

Our final step is to present an expression for the N$^2$MHV superamplitude in terms of the $s$ basis amplitudes obtained from summing over partitions $\lambda=[w_1,w_2,\ldots,w_{n-4}]$, and over all singlets $\cs_\lambda$ within each partition. We call an $SU(n-4)$ semi-standard Young tableau  a \emph{$\lambda$-compatible tableau} if it contains $w_1$ entries of index $1$, $w_2$ entries of index $2$, etc. For each $\lambda$, the number of $\lambda$-compatible tableaux is  $\cs_\lambda$. The superamplitude then takes the simple form
\begin{equation}
\begin{split}
   \cm^{\text{N$^2$MHV}}_n ~~=~&~\frac{1}{256}\sum_\lambda
   \frac{\mathcal{C}_\l}{(n\!-\!4)!}
     \sum_{{}^\text{$\lambda$-compatible}_\text{~~tableaux Y}}\!\!\! (-)^Y
     A_n\bigl(\,
     \bigl\{
    {}^{i_1 j_1 \cdots v_1}_{i_2 j_2 \cdots v_2}
    \bigr\}
     ++--\bigr)~
      Z^{i_1 j_1 \cdots v_1}_{i_2 j_2  \cdots v_2}\\[1ex]
     &~+~\text{momentum permutations}\,.
\end{split}
\end{equation}
Here, the sum goes over all permutations of the
 momentum labels 1, 2,$\,\ldots n\,$, and the combinatorial factor $\mathcal{C}_\l/(n-4)!$ compensates for the overcounting in this sum.\footnote{
In the notation of~(\ref{Cl}) in appendix~\ref{app17A}, we have
$(n\!-\!4)!/\mathcal{C}_\l=\prod_{w=0}^\cn t_w!$\,, \ie this combinatorial factor counts the permutations that leave the partition $\lambda$ invariant.
}
Note that precisely $s=\sum_\l\mathcal{S}_\l$ amplitudes go into the definition of this superamplitude, and thus no more than $s$ amplitudes need to be computed at any loop-level to fully determine the supergravity superamplitude $\cm_n$.


\subsection*{Acknowledgement}

We thank Zvi Bern and Renata Kallosh for the encouragement to write up these results. We are grateful to Alejandro Morales, Alexander Postnikov, and Richard Stanley for their generous help with the combinatorics of this project. We also thank Tim Cohen, Lance Dixon, and David Kosower for useful comments and suggestions.
HE is supported in part by the US Department of Energy under grant DE-FG02-95ER40899.
The research of DZF is supported by NSF grant
PHY-0600465 and by the US Department of Energy through cooperative research agreement DE-FG-0205FR41360.
MK is supported in part by NSF grant PHY-0756966.

\appendix

\setcounter{equation}{0}
\section{The general basis}\label{app18}
In Sections~\ref{secNMHV} and~\ref{secbeyond}  we solved the SUSY Ward identities, and presented expressions for superamplitudes in terms of a set of basis amplitudes in which we picked out the four external lines $n-3$, $n-2$, $n-1$, and $n$ as special. Clearly, this choice of lines was arbitrary and we could have picked out any other four lines
 $r$, $s$, $t$, and $u$. More generally, we could have picked a different set of four special lines $r_a$, $s_a$, $t_a$, $u_a$ for each $SU(\cn)$ index $a=1,\ldots,\cn$. The basis amplitudes then consist of all amplitudes that carry the $SU(\cn)$ index $a$ on lines $t_a$, $u_a$ and do {\em not} carry index $a$ on lines $r_a$ and $s_a$.
In terms of these basis amplitudes, the superamplitude at NMHV level takes the form
\be
   \ca_n^{\rm NMHV} ~=~
   \delta^{(2\cn)}\big( \tilde{Q}_a\big)\!\!\!
     \sum_{i_a\neq r_a,s_a,t_a,u_a}\!\!\!\!\!
     A_n\big(\{i_a,t_a,u_a\}\big)\,
     \prod_{a=1}^\cn
     \frac{\pdan_{i_a,r_a,s_a,a}}
     {[r_as_a]\<t_au_a\>}   \, .
\ee
Here, the basis amplitudes are defined as\footnote{
The right hand side is to be understood as follows. As usual, all external lines not mentioned are positive-helicity gluons. If any of the indices coincide, the corresponding
$SU(\cn)_R$
 indices are understood to merge onto the same line. For example, if $i_1=i_2$, we have $A_{i_1}^1\cdots A_{i_2}^2~\to~ A_{i_1}^{12}$. Similarly, if $i_{\cn}=t_1$, we have $A_{i_\cn}^\cn \cdots A_{t_1}^1~\to~A_{i\cn}^{\cn 1}=-A_{i_\cn}^{1\cn}$\,, etc. On the other hand, if the set $\{i_1,\ldots,i_\cn,t_1,\ldots,t_\cn,u_1,\ldots,u_\cn\}$ is not
  in arithmetic order,
 the definition on the right-hand side must be adjusted to reflect the correct line-ordering. For example, if $i_{\cn}>t_{1}$, we have $A_{i_\cn}^\cn \cdots A_{t_1}^1~\to~-A_{t_1}^1\cdots A_{i_\cn}^\cn$\,.}
\begin{equation}
    A_n\big(\{i_a,t_a,u_a\}\big)~=~
    \big\<\cdots A_{i_1}^1 \cdots  A_{i_\cn}^\cn \cdots A_{t_1}^1 \cdots  A_{t_\cn}^\cn \cdots A_{u_1}^1 \cdots  A_{u_\cn}^\cn\cdots\big\>\,.
\end{equation}
To recover the solution to the SUSY Ward identities of Section~\ref{secNMHV}, we set $r_a=n-3$, $s_a=n-2$, $t_a=n-1$, and $u_a=n$ for all $a$. Furthermore, we also change notation $\{i_1,i_2,\ldots,i_\cn\}\to\{i,j,k,l\}$ for $\cn=4$ SYM, and $\{i_1,i_2,\ldots,i_\cn\}\to\{i,j,k,l,p,q,u,v\}$ for supergravity.

It is now straightforward to generalize this basis to N$^K$MHV level. Again the general basis consists of all amplitudes that carry the SU($\cn$) index $a$ on lines $t_a$, $u_a$ and do {\em not} carry index $a$ on lines $r_a$ and $s_a$. Each SU($\cn$) index $a$ still needs to be distributed $K$ times among the remaining lines. We label these lines by $i_{a,A}$, $A=1,\ldots,K$. The superamplitude then reads
\be
   \ca_n^\text{N$^k$MHV} ~=~
   \delta^{(2\cn)}\big( \tilde{Q}_a\big)
   \!\!\!\sum_{i_{a,A}\neq r_a,s_a,t_a,u_a}\!\!\!\!\!
     A_n\big(\{i_{a,A},t_a,u_a\}\big)\,
     \prod_{a=1}^\cn\Biggl[\<t_au_a\>^{-1} \prod_{B=1}^K\frac{\pdan_{i_{a\!,A},r_a,s_a,a}}{[r_as_a]} \Biggr] \, ,
\ee
with
\begin{equation}
    A_n\big(\{i_{a,A},t_a,u_a\}\big)~=~
    \big\<\cdots A_{i_{1,1}}^1 \cdots  A_{i_{\cn,K}}^4 \cdots A_{t_1}^1 \cdots  A_{t_\cn}^4 \cdots A_{u_1}^1 \cdots  A_{u_\cn}^4\cdots\big\>\,.
\end{equation}
This is our most general solution to the SUSY Ward identities at N$^K$MHV level. To recover the solution of Section~\ref{secbeyond}, we set $r_a=n-3$, $s_a=n-2$, $t_a=n-1$, and $u_a=n$ for all $a$, and also change notation $\{i_{1,A},i_{2,A},\ldots,i_{\cn,A}\}\to\{i_A,j_A,k_A,l_A\}$ for $\cn=4$ SYM, and
$\{i_{1,A},i_{2,A},\ldots,i_{\cn,A}\}\to\{i_A,j_A,k_A,l_A,p_A,q_A,u_A,v_A\}$ for supergravity.


\section{Group theory and counting}
\lab{app17}

We review here some group theory needed in our analysis.
Recall that irreps of $SU(q)$ are in one-to-one correspondence with Young diagrams with at most $q$ rows.  The dimension $d_Y$ of the irrep is most easily calculated from the  hook rule formula \cite{hook}.
It is central in our work that  $d_Y$ is equal to  the number of semi-standard tableaux constructed from $Y$ using integers from the set $\{1, 2, \dots, q\}$. In the main text, $q$ will be identified as $n-4$, where $n$ is the number of external states of an amplitude.
We review semi-standard tableaux below.

\subsection{Young diagrams and semi-standard tableaux}
\lab{app17A}

A \emph{semi-standard tableau} is obtained from a Young diagram $Y$ by inserting the numbers from the set $\{1,2,\dots,q\}$ into $Y$ according to the rules that
1) numbers must increase weakly along rows, and
2) numbers must increase strictly going down columns.
If a tableau contains $w_1$ 1's, $w_2$ 2's etc, then it is said to have \emph{weight} $(w_1,w_2,\dots,w_q)$.

For example, the rectangular Young diagram with 2 rows and 4 columns has three semi-standard tableaux of weight $(3,2,1,1,1)$, namely
\be
  \lab{sst1}
 \begin{array}{l}
  \framebox[3.5mm][c]{\scriptsize 1}
  \framebox[3.5mm][c]{\scriptsize 1}
  \framebox[3.5mm][c]{\scriptsize 1}
  \framebox[3.5mm][c]{\scriptsize 2}\\[-2pt]
  \framebox[3.5mm][c]{\scriptsize 2}
  \framebox[3.5mm][c]{\scriptsize 3}
  \framebox[3.5mm][c]{\scriptsize 4}
  \framebox[3.5mm][c]{\scriptsize 5}
  \end{array},~~~
   \begin{array}{l}
  \framebox[3.5mm][c]{\scriptsize 1}
  \framebox[3.5mm][c]{\scriptsize 1}
  \framebox[3.5mm][c]{\scriptsize 1}
  \framebox[3.5mm][c]{\scriptsize 3}\\[-2pt]
  \framebox[3.5mm][c]{\scriptsize 2}
  \framebox[3.5mm][c]{\scriptsize 2}
  \framebox[3.5mm][c]{\scriptsize 4}
  \framebox[3.5mm][c]{\scriptsize 5}
  \end{array},~~~
   \begin{array}{l}
  \framebox[3.5mm][c]{\scriptsize 1}
  \framebox[3.5mm][c]{\scriptsize 1}
  \framebox[3.5mm][c]{\scriptsize 1}
  \framebox[3.5mm][c]{\scriptsize 4}\\[-2pt]
  \framebox[3.5mm][c]{\scriptsize 2}
  \framebox[3.5mm][c]{\scriptsize 2}
  \framebox[3.5mm][c]{\scriptsize 3}
  \framebox[3.5mm][c]{\scriptsize 5}
  \end{array}.
\ee
The multiplicities can be assigned in any order. For instance, the  three semi-standard tableaux with weight $(1,2,1,3,1)$ are
\be
  \lab{sst2}
 \begin{array}{l}
  \framebox[3.5mm][c]{\scriptsize 1}
  \framebox[3.5mm][c]{\scriptsize 2}
  \framebox[3.5mm][c]{\scriptsize 2}
  \framebox[3.5mm][c]{\scriptsize 3}\\[-2pt]
  \framebox[3.5mm][c]{\scriptsize 4}
  \framebox[3.5mm][c]{\scriptsize 4}
  \framebox[3.5mm][c]{\scriptsize 4}
  \framebox[3.5mm][c]{\scriptsize 5}
  \end{array},~~~
   \begin{array}{l}
  \framebox[3.5mm][c]{\scriptsize 1}
  \framebox[3.5mm][c]{\scriptsize 2}
  \framebox[3.5mm][c]{\scriptsize 2}
  \framebox[3.5mm][c]{\scriptsize 4}\\[-2pt]
  \framebox[3.5mm][c]{\scriptsize 3}
  \framebox[3.5mm][c]{\scriptsize 4}
  \framebox[3.5mm][c]{\scriptsize 4}
  \framebox[3.5mm][c]{\scriptsize 5}
  \end{array},~~~
   \begin{array}{l}
  \framebox[3.5mm][c]{\scriptsize 1}
  \framebox[3.5mm][c]{\scriptsize 2}
  \framebox[3.5mm][c]{\scriptsize 3}
  \framebox[3.5mm][c]{\scriptsize 4}\\[-2pt]
  \framebox[3.5mm][c]{\scriptsize 2}
  \framebox[3.5mm][c]{\scriptsize 4}
  \framebox[3.5mm][c]{\scriptsize 4}
  \framebox[3.5mm][c]{\scriptsize 5}
  \end{array}.
\ee

Let $Y$ be a given Young diagram with $p$ boxes.
If $\lambda = [w_1,w_2,\dots,w_q]$ is a partition of $p$, then any ordering of the $w_i$'s can be used as a weight
 that determines a set of semi-standard tableaux of $Y$. The number of distinct weights of a given partition $\lambda$ is simply the multinomial coefficient \cite{stanley}
\be
  \lab{Cl}
  \mathcal{C}_\lambda = \frac{q!}{\prod_{w=0}^\cn (t_w!)}\, ,
\ee
where $t_w$ is the
number of times the number $w$, with $0 \le w \le \cn$, occurs in the partition $\l$.
For example, $\l = [3,2,1,1,1]$ has $t_0=0$, $t_1=3$, $t_2=1$, and $t_3=1$, and gives
$\mathcal{C}_\lambda = 5!/3! = 20$.

The fact that we found three
 semi-standard tableaux in both \reef{sst1} and \reef{sst2} is no coincidence. Actually, each of the $\mathcal{C}_\lambda=20$ different weight assignments associated with the partition $\l = [3,2,1,1,1]$ gives three semi-standard tableaux.
The number of semi-standard tableaux with any weight from a partition $\lambda$ is the \emph{Kostka number} $\mathcal{S}_\lambda$. As seen in our example,
$\mathcal{S}_\lambda$ depends only on the partition $\lambda$, not on the particular weight \cite{fulton}.

For the given Young diagram $Y$ with $p$ boxes, the total number of semi-standard tableaux with weights from a partition $\lambda$ is $\mathcal{C}_\lambda \mathcal{S}_\lambda$. The total number of
semi-standard tableaux containing numbers from the set $\{1,2,\dots,q\}$ is $d_Y = \sum_{|\lambda|=q}\mathcal{C}_\lambda \mathcal{S}_\lambda$, where the sum is over all  partitions of $p$ of length $q$. This number is also the dimension of the irrep of $SU(q)$ corresponding to the Young diagram $Y$.

\subsection{The Kostka number $\mathcal{S}_\l$ counts singlets}

In our applications we are interested only in rectangular Young diagrams with $K$ rows and $\cn$ columns. The weights of the associated semi-standard tableaux are constructed from partitions $\lambda=[n_1,n_2,\dots,n_q]$ of the integer
 $p=\cn K$  such that the length of $\lambda$ is $q$ and $0 \le n_i \le \cn$.

For each number $n_i$ of the partition $\lambda=[n_1,n_2,\dots,n_q]$, let us associate the $n_i$-index fully antisymmetric irrep of $SU(\cn)$. The corresponding Young diagram $Y_{n_i}$ is a single column of length $n_i$. The cases $n_i=0$ and $n_i=\cn$ both represent the singlet, $n_i=1$ is the fundamental representation $\Box$, while $n_i=\cn-1$ is the anti-fundamental $\bar\Box$, etc. We now demonstrate that the Kostka number $\mathcal{S}_\l$ counts the number of singlets in the decomposition of the product $Y_{n_1} \otimes Y_{n_2} \otimes \cdots \otimes  Y_{n_q}$ in $SU(\cn)$.

The decomposition of a product of two irreps can be found by attaching the boxes of  the second Young diagram to the first one in all possible ways that do not violate the (anti)symmetrizations of the boxes in the second Young diagram. If the two original Young diagrams have $m_1$ and $m_2$ boxes, then each diagram in the decomposition has $m_1+m_2$ boxes. Also, in $SU(\cn)$ no column can be longer than $\cn$.
Singlets in the decomposition are rectangular diagrams with $\cn$ rows.

Each diagram in the decomposition of $Y_{n_1} \otimes Y_{n_2} \otimes \cdots \otimes  Y_{n_q}$ contains $\sum_{i=1}^q n_i = \cn K$ boxes, and a singlet is therefore a rectangular $\cn$-by-$K$ diagram. To keep track of the boxes from each diagram $Y_{n_i}$, let us put the number $i$ in all the boxes of the $i$th diagram $Y_{n_i}$. Consider an example with $\cn=4$, $K=2$ and $\lambda = [3,2,1,1,1]$. The product $Y_{n_1} \otimes Y_{n_2} \otimes \cdots \otimes  Y_{n_5}$ decomposes as follows:
\be
 \begin{array}{l}
  \framebox[3.5mm][c]{\scriptsize 1}\\[-2pt]
  \framebox[3.5mm][c]{\scriptsize 1}\\[-2pt]
  \framebox[3.5mm][c]{\scriptsize 1}
  \end{array}
  \otimes
  \begin{array}{l}
  \framebox[3.5mm][c]{\scriptsize 2}\\[-2pt]
  \framebox[3.5mm][c]{\scriptsize 2}
  \end{array}
  \otimes
   \begin{array}{l}
  \framebox[3.5mm][c]{\scriptsize 3}
  \end{array}
  \otimes
   \begin{array}{l}
  \framebox[3.5mm][c]{\scriptsize 4}
  \end{array}
  \otimes
   \begin{array}{l}
  \framebox[3.5mm][c]{\scriptsize 5}
  \end{array}
   ~=~
  \begin{array}{l}
  \framebox[3.5mm][c]{\scriptsize 1}
  \framebox[3.5mm][c]{\scriptsize 2}\\[-2pt]
  \framebox[3.5mm][c]{\scriptsize 1}
  \framebox[3.5mm][c]{\scriptsize 3}\\[-2pt]
  \framebox[3.5mm][c]{\scriptsize 1}
  \framebox[3.5mm][c]{\scriptsize 4}\\[-2pt]
  \framebox[3.5mm][c]{\scriptsize 2}
  \framebox[3.5mm][c]{\scriptsize 5}
  \end{array}
  +
  \begin{array}{l}
  \framebox[3.5mm][c]{\scriptsize 1}
  \framebox[3.5mm][c]{\scriptsize 2}\\[-2pt]
  \framebox[3.5mm][c]{\scriptsize 1}
  \framebox[3.5mm][c]{\scriptsize 2}\\[-2pt]
  \framebox[3.5mm][c]{\scriptsize 1}
  \framebox[3.5mm][c]{\scriptsize 4}\\[-2pt]
  \framebox[3.5mm][c]{\scriptsize 3}
  \framebox[3.5mm][c]{\scriptsize 5}
  \end{array}
  +
   \begin{array}{l}
  \framebox[3.5mm][c]{\scriptsize 1}
  \framebox[3.5mm][c]{\scriptsize 2}\\[-2pt]
  \framebox[3.5mm][c]{\scriptsize 1}
  \framebox[3.5mm][c]{\scriptsize 2}\\[-2pt]
  \framebox[3.5mm][c]{\scriptsize 1}
  \framebox[3.5mm][c]{\scriptsize 3}\\[-2pt]
  \framebox[3.5mm][c]{\scriptsize 4}
  \framebox[3.5mm][c]{\scriptsize 5}
  \end{array}
  +
  \text{non-singlets}.
  \lab{decomp}
\ee
The conjugate (obtained by reflection along the diagonal) of the first singlet diagram gives
\be
  \lab{sing1}
  \begin{array}{l}
  \framebox[3.5mm][c]{\scriptsize 1}
  \framebox[3.5mm][c]{\scriptsize 2}\\[-2pt]
  \framebox[3.5mm][c]{\scriptsize 1}
  \framebox[3.5mm][c]{\scriptsize 3}\\[-2pt]
  \framebox[3.5mm][c]{\scriptsize 1}
  \framebox[3.5mm][c]{\scriptsize 4}\\[-2pt]
  \framebox[3.5mm][c]{\scriptsize 2}
  \framebox[3.5mm][c]{\scriptsize 5}
  \end{array}
  ~\hookrightarrow
  \begin{array}{l}
  \framebox[3.5mm][c]{\scriptsize 1}
  \framebox[3.5mm][c]{\scriptsize 1}
  \framebox[3.5mm][c]{\scriptsize 1}
  \framebox[3.5mm][c]{\scriptsize 2}\\[-2pt]
  \framebox[3.5mm][c]{\scriptsize 2}
  \framebox[3.5mm][c]{\scriptsize 3}
  \framebox[3.5mm][c]{\scriptsize 4}
  \framebox[3.5mm][c]{\scriptsize 5}
  \end{array}.
\ee
The new tableau is precisely the first semi-standard tableau of \reef{sst1}.
The two other semi-standard tableaux of \reef{sst1} are found by conjugating the two other singlets in \reef{decomp}. Moreover, taking the product in any other order does not change the number of singlets in the decomposition, but the semi-standard tableaux are different. For instance, in the product
\begin{equation}
\begin{array}{l}
  \framebox[3.5mm][c]{\scriptsize 1}
  \end{array}
  \otimes
  \begin{array}{l}
  \framebox[3.5mm][c]{\scriptsize 2}\\[-2pt]
  \framebox[3.5mm][c]{\scriptsize 2}
  \end{array}
  \otimes
   \begin{array}{l}
  \framebox[3.5mm][c]{\scriptsize 3}
  \end{array}
  \otimes
  \begin{array}{l}
    \framebox[3.5mm][c]{\scriptsize 4}\\[-2pt]
  \framebox[3.5mm][c]{\scriptsize 4}\\[-2pt]
  \framebox[3.5mm][c]{\scriptsize 4}
  \end{array}
  \otimes
  \begin{array}{l}
  \framebox[3.5mm][c]{\scriptsize 5}
  \end{array}\,,
\end{equation}
the three singlets reproduce, after conjugation, exactly the semi-standard tableaux of \reef{sst2}. The number of different ways we can take the product is given by the multinomial coefficient $\mathcal{C}_\l = 5!/(3!\, 2!) = 20$. The decomposition of each product contains precisely three singlets, $\mathcal{S}_\lambda = 3$. Upon conjugation, each one of these will be a distinct semi-standard tableau. Thus, associated with the partition $\lambda = [3,2,1,1,1]$, we find $\mathcal{C}_\l \mathcal{S}_\l = 60$ distinct semi-standard tableaux. This is exactly right.

The example should suffice to convince the reader that the argument carries over to any partitions $\lambda$ of $\cn K$.
The fact that the number of singlets in the decomposition does not depend on the order of the product shows that the Kostka number only depends on the partition, and not on the weights.

We have shown that the Kostka numbers $\mathcal{S}_\l$ for rectangular $\cn$-by-$K$ Young diagrams have two roles. For a given partition $\lambda=[n_1,n_2,\dots,n_q]$ of $\cn K$
with $0 \le n_i \le \cn$ they count
1) the number of semi-standard tableaux with weight $\lambda$, and 2) singlets in the product  $Y_{n_1} \otimes Y_{n_2} \otimes \cdots \otimes  Y_{n_q}$ of  $SU(\cn)$ $n_i$-index fully antisymmetric irreps.
Both characterizations are useful in our analysis.



\begin{thebibliography}{99}

\bibitem{gris}
  M.~T.~Grisaru and H.~N.~Pendleton,
  ``Some Properties Of Scattering Amplitudes In Supersymmetric Theories,''
  Nucl.\ Phys.\  B {\bf 124}, 81 (1977).

\bibitem{gpvanN}
  M.~T.~Grisaru, H.~N.~Pendleton and P.~van Nieuwenhuizen,
  ``Supergravity And The S Matrix,''
  Phys.\ Rev.\  D {\bf 15}, 996 (1977).

\bibitem{PTsusy}
  S.~J.~Parke and T.~R.~Taylor,
  ``Perturbative QCD Utilizing Extended Supersymmetry,''
  Phys.\ Lett.\  B {\bf 157}, 81 (1985)
  [Erratum-ibid.\  {\bf 174B}, 465 (1986)].


\bibitem{Drummond:2006rz}
  J.~M.~Drummond, J.~Henn, V.~A.~Smirnov and E.~Sokatchev,
  ``Magic identities for conformal four-point integrals,''
  JHEP {\bf 0701}, 064 (2007)
  [arXiv:hep-th/0607160].

\bibitem{Alday:2007hr}
  L.~F.~Alday and J.~M.~Maldacena,
  ``Gluon scattering amplitudes at strong coupling,''
  JHEP {\bf 0706}, 064 (2007)
  [arXiv:0705.0303 [hep-th]].

\bibitem{Alday:2007he}
  L.~F.~Alday and J.~Maldacena,
  ``Comments on gluon scattering amplitudes via AdS/CFT,''
  JHEP {\bf 0711}, 068 (2007)
  [arXiv:0710.1060 [hep-th]].


\bibitem{Drummond:2008vq}
  J.~M.~Drummond, J.~Henn, G.~P.~Korchemsky and E.~Sokatchev,
  ``Dual superconformal symmetry of scattering amplitudes in N=4
  super-Yang-Mills theory,''
  arXiv:0807.1095 [hep-th].

\bibitem{Brandhuber:2008pf}
  A.~Brandhuber, P.~Heslop and G.~Travaglini,
  ``A note on dual superconformal symmetry of the N=4 super Yang-Mills S-matrix,''
  Phys.\ Rev.\  D {\bf 78}, 125005 (2008)
  [arXiv:0807.4097 [hep-th]].

\bibitem{DH}
  J.~M.~Drummond and J.~M.~Henn,
  ``All tree-level amplitudes in N=4 SYM,''
  JHEP {\bf 0904}, 018 (2009)
  [arXiv:0808.2475 [hep-th]].

\bibitem{Drummond:2008bq}
  J.~M.~Drummond, J.~Henn, G.~P.~Korchemsky and E.~Sokatchev,
  ``Generalized unitarity for N=4 super-amplitudes,''
  arXiv:0808.0491 [hep-th].

\bibitem{Drummond:2009fd}
  J.~M.~Drummond, J.~M.~Henn and J.~Plefka,
  ``Yangian symmetry of scattering amplitudes in N=4 super Yang-Mills theory,''
  JHEP {\bf 0905}, 046 (2009)
  [arXiv:0902.2987 [hep-th]].

\bibitem{Brandhuber:2009kh}
  A.~Brandhuber, P.~Heslop and G.~Travaglini,
  ``Proof of the Dual Conformal Anomaly of One-Loop Amplitudes in N=4 SYM,''
  JHEP {\bf 0910}, 063 (2009)
  [arXiv:0906.3552 [hep-th]].

\bibitem{Brandhuber:2009xz}
  A.~Brandhuber, P.~Heslop and G.~Travaglini,
  ``One-Loop Amplitudes in N=4 Super Yang-Mills and Anomalous Dual Conformal Symmetry,''
  JHEP {\bf 0908}, 095 (2009)
  [arXiv:0905.4377 [hep-th]].

\bibitem{Elvang:2009ya}
  H.~Elvang, D.~Z.~Freedman and M.~Kiermaier,
  ``Dual conformal symmetry of 1-loop NMHV amplitudes in N=4 SYM theory,''
  arXiv:0905.4379 [hep-th].

\bibitem{Bargheer:2009qu}
  T.~Bargheer, N.~Beisert, W.~Galleas, F.~Loebbert and T.~McLoughlin,
  ``Exacting N=4 Superconformal Symmetry,''
  arXiv:0905.3738 [hep-th].

\bibitem{Korchemsky:2009hm}
  G.~P.~Korchemsky and E.~Sokatchev,
  ``Symmetries and analytic properties of scattering amplitudes in N=4 SYM theory,''
  arXiv:0906.1737 [hep-th].

\bibitem{nair}
  V.~P.~Nair,
  ``A current algebra for some gauge theory amplitudes,''
  Phys.\ Lett.\  B {\bf 214}, 215 (1988).

\bibitem{witten}
  E.~Witten,
  ``Perturbative gauge theory as a string theory in twistor space,''
  Commun.\ Math.\ Phys.\  {\bf 252}, 189 (2004)
  [arXiv:hep-th/0312171].

\bibitem{khoze}
  G.~Georgiou, E.~W.~N.~Glover and V.~V.~Khoze,
  ``Non-MHV Tree Amplitudes in Gauge Theory,''
  JHEP {\bf 0407}, 048 (2004)
  [arXiv:hep-th/0407027].

  G.~Georgiou and V.~V.~Khoze,
  ``Tree amplitudes in gauge theory as scalar MHV diagrams,''
  JHEP {\bf 0405}, 070 (2004)
  [arXiv:hep-th/0404072].

\bibitem{BEF}
  M.~Bianchi, H.~Elvang and D.~Z.~Freedman,
  ``Generating Tree Amplitudes in N=4 SYM and N = 8 SG,''
  JHEP {\bf 0809}, 063 (2008)
  [arXiv:0805.0757 [hep-th]].

\bibitem{ArkaniHamed:2008gz}
  N.~Arkani-Hamed, F.~Cachazo and J.~Kaplan,
  ``What is the Simplest Quantum Field Theory?,''
  arXiv:0808.1446 [hep-th].





\bibitem{EFK1}
  H.~Elvang, D.~Z.~Freedman and M.~Kiermaier,
  ``Recursion Relations, Generating Functions, and Unitarity Sums in N=4 SYM Theory,''
  JHEP {\bf 0904}, 009 (2009)
  [arXiv:0808.1720 [hep-th]].

\bibitem{EFK2}
  H.~Elvang, D.~Z.~Freedman and M.~Kiermaier,
  ``Proof of the MHV vertex expansion for all tree amplitudes in N=4 SYM theory,''
  JHEP {\bf 0906}, 068 (2009)
  [arXiv:0811.3624 [hep-th]].


\bibitem{KN}
  M.~Kiermaier and S.~G.~Naculich,
  ``A super MHV vertex expansion for N=4 SYM theory,''
  JHEP {\bf 0905}, 072 (2009)
  [arXiv:0903.0377 [hep-th]].

\bibitem{Drummond:2009ge}
  J.~M.~Drummond, M.~Spradlin, A.~Volovich and C.~Wen,
  ``Tree-Level Amplitudes in N=8 Supergravity,''
  Phys.\ Rev.\  D {\bf 79}, 105018 (2009)
  [arXiv:0901.2363 [hep-th]].


\bibitem{ArkaniHamed:2009dn}
  N.~Arkani-Hamed, F.~Cachazo, C.~Cheung and J.~Kaplan,
  ``A Duality For The S Matrix,''
  arXiv:0907.5418 [hep-th].


\bibitem{linkrefs}
  M.~Spradlin and A.~Volovich,
  ``From Twistor String Theory To Recursion Relations,''
  Phys.\ Rev.\  D {\bf 80}, 085022 (2009)
  [arXiv:0909.0229 [hep-th]].

  L.~Mason and D.~Skinner,
  ``Dual Superconformal Invariance, Momentum Twistors and Grassmannians,''
  arXiv:0909.0250 [hep-th].

  N.~Arkani-Hamed, F.~Cachazo and C.~Cheung,
  ``The Grassmannian Origin Of Dual Superconformal Invariance,''
  arXiv:0909.0483 [hep-th].

  L.~Dolan and P.~Goddard,
  ``Gluon Tree Amplitudes in Open Twistor String Theory,''
  arXiv:0909.0499 [hep-th].



\bibitem{NimaParis}
N.~Arkani-Hamed, ``{What is the Simplest QFT?},''
{talk given at the Paris Workshop
{\em Wonders of Gauge Theory and Supergravity}},
{June 24, 2008}.


\bibitem{kunszt}
Z.~Kunszt,
"Combined use of the CALKUL method and N = 1 supersymmetry to calculate QCD six-parton processes,"
Nucl.~Phys.~B 271, 333 (1986).

\bibitem{Bidder:2005in}
  S.~J.~Bidder, D.~C.~Dunbar and W.~B.~Perkins,
  ``Supersymmetric Ward identities and NMHV amplitudes involving gluinos,''
  JHEP {\bf 0508}, 055 (2005)
  [arXiv:hep-th/0505249].


\bibitem{Berends:1987me}
  F.~A.~Berends and W.~T.~Giele,
  ``Recursive Calculations for Processes with n Gluons,''
  Nucl.\ Phys.\  B {\bf 306}, 759 (1988).

\bibitem{Kleiss:1988ne}
  R.~Kleiss and H.~Kuijf,
  ``Multi-Gluon Cross-Sections and Five Jet Production at Hadron Colliders,''
  Nucl.\ Phys.\  B {\bf 312}, 616 (1989).


\bibitem{Bern:2008qj}
  Z.~Bern, J.~J.~M.~Carrasco and H.~Johansson,
  ``New Relations for Gauge-Theory Amplitudes,''
  Phys.\ Rev.\  D {\bf 78}, 085011 (2008)
  [arXiv:0805.3993 [hep-ph]].

\bibitem{BjerrumBohr:2009rd}
  N.~E.~J.~Bjerrum-Bohr, P.~H.~Damgaard and P.~Vanhove,
  ``Minimal Basis for Gauge Theory Amplitudes,''
  Phys.\ Rev.\ Lett.\  {\bf 103}, 161602 (2009)
  [arXiv:0907.1425 [hep-th]].

\bibitem{Stieberger:2009hq}
  S.~Stieberger,
  ``Open and Closed vs. Pure Open String Disk Amplitudes,''
  arXiv:0907.2211 [hep-th].





\bibitem{ST2007}
  S.~Stieberger and T.~R.~Taylor,
  ``Supersymmetry Relations and MHV Amplitudes in Superstring Theory,''
  Nucl.\ Phys.\  B {\bf 793}, 83 (2008)
  [arXiv:0708.0574 [hep-th]].

\bibitem{Stieberger2009}
  S.~Stieberger,
  ``On tree-level higher order gravitational couplings in superstring theory,''
  arXiv:0910.0180 [hep-th].

\bibitem{Obers2008}
  R.~Boels, K.~J.~Larsen, N.~A.~Obers and M.~Vonk,
  ``MHV, CSW and BCFW: field theory structures in string theory amplitudes,''
  JHEP {\bf 0811}, 015 (2008)
  [arXiv:0808.2598 [hep-th]].

\bibitem{N4loop}
See for example,
  Z.~Bern, L.~J.~Dixon, D.~A.~Kosower, R.~Roiban, M.~Spradlin, C.~Vergu and A.~Volovich,
  ``The Two-Loop Six-Gluon MHV Amplitude in Maximally Supersymmetric Yang-Mills Theory,''
  Phys.\ Rev.\  D {\bf 78}, 045007 (2008)
  [arXiv:0803.1465 [hep-th]].

  J.~M.~Drummond, J.~Henn, G.~P.~Korchemsky and E.~Sokatchev,
  ``Hexagon Wilson loop = six-gluon MHV amplitude,''
  Nucl.\ Phys.\  B {\bf 815}, 142 (2009)
  [arXiv:0803.1466 [hep-th]].


\bibitem{Bern:2007hh}
  Z.~Bern, J.~J.~Carrasco, L.~J.~Dixon, H.~Johansson, D.~A.~Kosower and R.~Roiban,
  ``Three-Loop Superfiniteness of N=8 Supergravity,''
  Phys.\ Rev.\ Lett.\  {\bf 98}, 161303 (2007)
  [arXiv:hep-th/0702112].

\bibitem{Bern:2008pv}
  Z.~Bern, J.~J.~M.~Carrasco, L.~J.~Dixon, H.~Johansson and R.~Roiban,
  ``Manifest Ultraviolet Behavior for the Three-Loop Four-Point Amplitude of N=8 Supergravity,''
  Phys.\ Rev.\  D {\bf 78}, 105019 (2008)
  [arXiv:0808.4112 [hep-th]].

\bibitem{Bern:2009kd}
  Z.~Bern, J.~J.~Carrasco, L.~J.~Dixon, H.~Johansson and R.~Roiban,
  ``The Ultraviolet Behavior of N=8 Supergravity at Four Loops,''
  Phys.\ Rev.\ Lett.\  {\bf 103}, 081301 (2009)
  [arXiv:0905.2326 [hep-th]].


\bibitem{susyCTs}
  R.~E.~Kallosh,
  ``Counterterms in extended supergravities,''
  Phys.\ Lett.\  B {\bf 99}, 122 (1981).



  P.~S.~Howe, K.~S.~Stelle and P.~K.~Townsend,
  ``Miraculous Ultraviolet Cancellations In Supersymmetry Made Manifest,''
  Nucl.\ Phys.\  B {\bf 236}, 125 (1984).

  P.~S.~Howe, K.~S.~Stelle and P.~K.~Townsend,
  ``Superactions,''
  Nucl.\ Phys.\  B {\bf 191}, 445 (1981).

  J.~M.~Drummond, P.~J.~Heslop, P.~S.~Howe and S.~F.~Kerstan,
  ``Integral invariants in N = 4 SYM and the effective action for  coincident
  D-branes,''
  JHEP {\bf 0308}, 016 (2003)

  R.~Kallosh,
  ``On UV Finiteness of the Four Loop N=8 Supergravity,''
  JHEP {\bf 0909}, 116 (2009)
  [arXiv:0906.3495 [hep-th]].

  G.~Bossard, P.~S.~Howe and K.~S.~Stelle,
  ``A note on the UV behaviour of maximally supersymmetric Yang-Mills theories,''
  arXiv:0908.3883 [hep-th].

  G.~Bossard, P.~S.~Howe and K.~S.~Stelle,
  ``The ultra-violet question in maximally supersymmetric field theories,''
  Gen.\ Rel.\ Grav.\  {\bf 41}, 919 (2009)
  [arXiv:0901.4661 [hep-th]].

\bibitem{Bern:2005iz}
  Z.~Bern, L.~J.~Dixon and V.~A.~Smirnov,
   ``Iteration of planar amplitudes in maximally supersymmetric Yang-Mills
  theory at three loops and beyond,''
  Phys.\ Rev.\  D {\bf 72}, 085001 (2005)
  [arXiv:hep-th/0505205].

\bibitem{Anastasiou:2003kj}
  C.~Anastasiou, Z.~Bern, L.~J.~Dixon and D.~A.~Kosower,
  ``Planar amplitudes in maximally supersymmetric Yang-Mills theory,''
  Phys.\ Rev.\ Lett.\  {\bf 91}, 251602 (2003)
  [arXiv:hep-th/0309040].



\bibitem{Drummond:2007bm}
  J.~M.~Drummond, J.~Henn, G.~P.~Korchemsky and E.~Sokatchev,
  ``The hexagon Wilson loop and the BDS ansatz for the six-gluon amplitude,''
  Phys.\ Lett.\  B {\bf 662}, 456 (2008)
  [arXiv:0712.4138 [hep-th]].


\bibitem{Anastasiou:2009kna}
  C.~Anastasiou, A.~Brandhuber, P.~Heslop, V.~V.~Khoze, B.~Spence and G.~Travaglini,
  ``Two-Loop Polygon Wilson Loops in N=4 SYM,''
  JHEP {\bf 0905}, 115 (2009)
  [arXiv:0902.2245 [hep-th]].

\bibitem{Bern:2009xq}
  Z.~Bern, J.~J.~M.~Carrasco, H.~Ita, H.~Johansson and R.~Roiban,
  ``On the Structure of Supersymmetric Sums in Multi-Loop Unitarity Cuts,''
  Phys.\ Rev.\  D {\bf 80}, 065029 (2009)
  [arXiv:0903.5348 [hep-th]].


\bibitem{hook}
\emph{The hook rule formula for the dimension of the irreps of $SU(n)$ can be found in}
H. Georgi, Lie Algebras in Particle Physics (Perseus Books, Reading, MA, 1999).

\emph{A diagrammatic proof was given in}
  H.~Elvang, P.~Cvitanovic and A.~D.~Kennedy,
  ``Diagrammatic Young Projection Operators for U(n),''
  J.\ Math.\ Phys.\ {\bf 46}, 043501 (2005)
  [arXiv:hep-th/0307186].

\bibitem{stanley}
R.~E.~Stanley,   "Enumerative Combinatorics. Vol I,"
(Cambridge University Press, 1997).

\bibitem{fulton}
W.~Fulton,   "Young tableaux: with applications to representation theory and geometry,"
(Cambridge University Press, 1997).


\bibitem{maccy} http://www.math.uiuc.edu/Macaulay2/

\end{thebibliography}
\end{document}